\documentclass{emulateapj}
\usepackage{apjfonts}

\usepackage{lscape}
\usepackage{graphicx}
\usepackage{natbib}
\usepackage{longtable}
\newcommand{\hbeta}{H{$\beta$}}
\newcommand{\halpha}{H{$\alpha$}}
\newcommand{\CIV}{C{\sevenrm IV}}

\newcommand{\MgII}{Mg{\sevenrm II}}
\newcommand{\OIII}{[O{\sevenrm\,III}]}
\newcommand{\OIIIa}{[O{\sevenrm\,III}]\,$\lambda$4959}
\newcommand{\OIIIb}{[O{\sevenrm\,III}]\,$\lambda$5007}
\newcommand{\OIIIab}{[O{\sevenrm\,III}]\,$\lambda\lambda$4959,5007}
\newcommand{\NII}{[N{\sevenrm\,II}]}

\newcommand{\NIIb}{[N{\sevenrm\,II}]\,$\lambda$6584}
\newcommand{\NIIab}{[N{\sevenrm\,II}]\,$\lambda\lambda$6548,6584}
\newcommand{\SII}{[S{\sevenrm\,II}]}
\newcommand{\SIIa}{[S{\sevenrm\,II}]\,$\lambda$6717}
\newcommand{\SIIb}{[S{\sevenrm\,II}]\,$\lambda$6731}
\newcommand{\SIIab}{[S{\sevenrm\,II}]\,$\lambda\lambda$6717,6731}

 \font\sevenrm=cmr7 scaled 1000

\begin{document}

\title{A Catalog of Quasar Properties from SDSS DR7}

\shorttitle{PROPERTIES OF SDSS DR7 QUASARS}

%\slugcomment{Draft Version}

\shortauthors{SHEN ET AL.}
\author{Yue Shen\altaffilmark{1}, Gordon T. Richards\altaffilmark{2}, Michael A. Strauss\altaffilmark{3},
Patrick B. Hall\altaffilmark{4}, Donald P. Schneider\altaffilmark{5},
 Stephanie Snedden\altaffilmark{6}, Dmitry Bizyaev\altaffilmark{6},
Howard Brewington\altaffilmark{6}, Viktor Malanushenko\altaffilmark{6},
Elena Malanushenko\altaffilmark{6}, Dan Oravetz\altaffilmark{6}, Kaike Pan\altaffilmark{6},
Audrey Simmons\altaffilmark{6}}
%\altaffiltext{1}{Princeton University Observatory, Princeton, NJ 08544.}
\altaffiltext{1}{Harvard-Smithsonian Center for Astrophysics, 60
Garden St., MS-51, Cambridge, MA 02138, USA.}
\altaffiltext{2}{Department of Physics, Drexel University, 3141
Chestnut Street, Philadelphia, PA 19104, USA.}
\altaffiltext{3}{Princeton University Observatory, Princeton, NJ
08544, USA.}
\altaffiltext{4}{Dept. of Physics \& Astronomy, York University,
4700 Keele St., Toronto, ON, M3J 1P3, Canada.}
\altaffiltext{5}{Department of Astronomy and Astrophysics, 525
Davey Laboratory, Pennsylvania State University, University Park,
PA 16802, USA.}
\altaffiltext{6}{Apache Point Observatory, Sunspot, NM, 88349, USA.}

\begin{abstract}
We present a compilation of properties of the 105,783 quasars
in the SDSS Data Release 7 (DR7) quasar catalog. In this
product, we compile continuum and emission line measurements
around the \halpha, \hbeta, \MgII\ and \CIV\ regions, as well
as other quantities such as radio properties, and flags
indicating broad absorption line quasars (BALQSOs), disk
emitters, etc. We also compile virial black hole mass estimates
based on various calibrations. For the fiducial virial mass
estimates we use the Vestergaard \& Peterson (VP06)
calibrations for \hbeta\ and \CIV, and our own calibration for
\MgII\ which matches the VP06 \hbeta\ masses on average. We
describe the construction of this catalog, and discuss its
limitations. The catalog and its future updates will be made
publicly available online.
%These measurements can be used to estimate virial black hole (BH)
%masses for $\sim 100,000$ SDSS quasars. We describe the catalog
%format and discuss correlations among spectral properties.
\end{abstract}

\section{Introduction}\label{sec:intro}
In recent years, studies of quasars and active galactic
nuclei (AGNs) have been greatly facilitated by dedicated
large-scale wide and deep field surveys in different bands, most
notably by optical surveys such as the Sloan Digital Sky Survey
\citep[SDSS,][]{SDSS} and the 2QZ survey
\citep[][]{Croom_etal_2004}. Indeed, the growing body of data has
revolutionized the study of quasars and AGNs.
Large, homogeneous data sets allow detailed investigations of the
phenomenological properties of quasars and AGNs, offering new
insights into the central engine powering these objects and their
connections to their host galaxies, especially when combined with
multi-wavelength coverage.
%\citep[e.g.,][]{Sulentic_etal_2000b,Croom_etal_2002,McLure_Dunlop_2004,Fine_etal_2006,Kollmeier_etal_2006,
%Scoville_etal_2007a,Shen_etal_2008b, Hickox_etal_2009}.
At the same time, it has become important to fit the
quasar/AGN population into its cosmological context, i.e., how
the supermassive black hole (SMBH) population evolves
across cosmic time. These data have led to a coherent picture of the cosmic evolution of the SMBH population within the concordance $\Lambda$CDM
paradigm \citep[][]{Kauffmann_Haehnelt_2000, Wyithe_Loeb_2003,Hopkins_etal_2006,Hopkins_etal_2008a,Shankar_etal_2009a,Shen_2009},
where the key observational components include
quasar clustering, the luminosity function (LF), the BH mass function, and the correlations between BHs and their host properties. Increasingly
larger
data sets are offering unique opportunities to
measure these properties with unprecedented precision.

In an earlier study of the virial BH mass and Eddington
ratio distributions of quasars, we measured spectral
properties for the SDSS Data Release 5 (DR5) quasar catalog
\citep{Schneider_etal_2007,Shen_etal_2008b}. We hereby extend this
exercise to the DR7 quasar catalog \citep{Schneider_etal_2010}. We now
include a more complete compilation than before of quantities from our
spectral fits. Our measurements are more sophisticated than the SDSS pipeline outputs in many ways, and are hence of
practical value. We describe the parent quasar sample in
\S\ref{sec:sample}, and the spectral measurements and the catalog
format in \S\ref{sec:spec_measure}. We discuss possible
applications of our measurements in \S\ref{sec:app}. Throughout
this paper we use cosmological parameters $\Omega_\Lambda=0.7$,
$\Omega_0=0.3$ and $h=0.7$.

\section{The Sample}\label{sec:sample}
The SDSS uses a dedicated 2.5-m wide-field telescope
\citep{Gunn_etal_2006} with a drift-scan camera with 30 $2048
\times 2048$ CCDs \citep{Gunn_etal_1998} to image the sky in five
broad bands \citep[$u\,g\,r\,i\,z$;][]{Fukugita_etal_1996}.  The
imaging data are taken on dark photometric nights of good seeing
\citep{Hogg_etal_2001}, are calibrated photometrically
\citep{Smith_etal_2002, Ivezic_etal_2004, Tucker_etal_2006} and
astrometrically \citep{Pier_etal_2003}, and object parameters are
measured \citep{Lupton_etal_2001, SDSS_EDR}. Quasar candidates
\citep{Richards_etal_2002a} for follow-up spectroscopy are
selected from the imaging data using their colors, and are
arranged in spectroscopic plates \citep{Blanton_etal_2003} to be
observed with a pair of fiber-fed double spectrographs.

\begin{figure}
  \centering
    \includegraphics[width=0.45\textwidth]{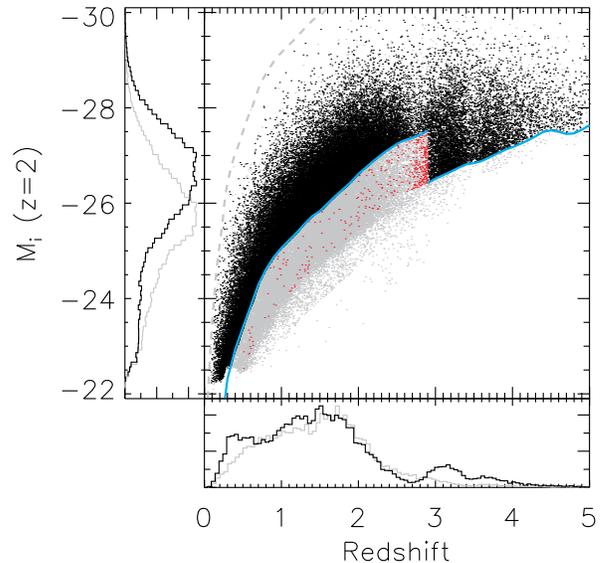}
    \caption{The distribution of DR7 quasars in luminosity-redshift space, where the left and bottom panels show the luminosity and redshift
    histograms. Luminosity is indicated using the (continuum and emission line) $K$-corrected, $i$-band absolute magnitude, $M_{i}(z=2)$, normalized
    at $z=2$ (Richards et~al. 2006a). The black dots are uniformly-selected
    quasars (see Richards et al.\ 2002a and \S\ref{subsec:cat}) and the gray dots are quasars selected by a variety of earlier algorithms or
    serendipitous selections. The red dots
    are selected by the {\tt QSO\_HiZ} uniform selection (e.g., Richards et~al. 2002a) but with $i>19.1$ and at $z<2.9$,  and should be removed in
    constructing
    homogeneous quasar samples. The cyan lines show the corresponding $M_i(z=2)$ for $i=19.1$ ($z<2.9$)
    and $i=20.2$ ($z>2.9$) respectively, and the gray dashed line shows the equivalent for $i=15$ (the bright limit for SDSS quasar targets). The
    non-uniformly selected quasars (gray dots) are targeted to fainter luminosities than are the uniformly selected quasars.}
    \label{fig:mi_dist}
\end{figure}

Our parent sample is the latest compilation of the spectroscopic
quasar catalog \citep{Schneider_etal_2010} from SDSS DR7 \citep[][]{SDSS_DR7}.
This sample contains 105,783 bona
fide quasars brighter than $M_{i}=-22.0$ and have at
least one broad emission line with FWHM larger than 1000 ${\rm
 km\ s^{-1}}$ or have interesting/complex absorption features. About
half of these objects are selected uniformly using the final
quasar target selection algorithm described in
\citet{Richards_etal_2002a}, with the remaining objects selected
via early versions of the target selection algorithm or various serendipitous
algorithms \citep[see][]{Schneider_etal_2010}, whose selection
completeness cannot be readily quantified. For statistical studies
such as quasar clustering and the LF, one should use the uniformly selected
quasar sample. %For the DR7 uniform quasar sample, the sky coverage
%is $\sim 6250\ {\rm deg}^2$.
Fig. \ref{fig:mi_dist} shows the distribution
of the 105,783 quasars in the redshift-luminosity plane.

%produced by the Princeton pipeline(Schlegel \etal, in preparation), which are the {\tt specBS} outputs
The reduced one-dimensional (1D) spectral data used in this study are available through the SDSS Data Archive
Server\footnote{http://das.sdss.org/spectro/} (DAS). The spectral resolution is
$R\sim 1850-2200$, and the 1D spectra are stored in vacuum
wavelength, with a pixel scale of $10^{-4}$ in log-wavelength, which
corresponds to $\sim 69\ {\rm kms^{-1}}$. Since the 6th SDSS data release
\citep[DR6,][]{SDSS_DR6}, the spectral flux calibration is scaled to the point spread function
(PSF) magnitudes of standard stars, therefore there is no longer need for a fiber-to-PSF conversion for the spectral flux (Shen et al.\ 2008 also
used
the PSF spectral flux calibration). Throughout the paper, we refer to the signal-to-noise ratio per pixel as S/N.

To include radio properties, we match the DR7 quasar catalog with
the FIRST \citep{White_etal_1997} catalog\footnote{The version of the FIRST source catalog
and the coverage maps used are as of July 16, 2008
(http://sundog.stsci.edu/first/catalogs/readme.html).} with a matching radius of 30"
 and estimate the radio loudness $R=f_{\rm
6cm}/f_{2500}$ following \citet{Jiang_etal_2007a}, where $f_{\rm
6cm}$ and $f_{2500}$ are the flux density ($f_\nu$) at rest-frame
6 cm and $2500$ \AA, respectively. For the quasars that have only one FIRST source within 30" we match them again to the FIRST catalog with a
matching
radius 5" and classify the matched ones as core-dominant radio quasars. Those quasars that have multiple FIRST source matches within 30" are
classified as lobe-dominated. The rest-frame 6 cm flux density is
determined from the FIRST integrated flux density at 20 cm
assuming a power-law slope of $\alpha_\nu=-0.5$; the rest-frame
$2500$ \AA\ flux density is determined from the power-law
continuum fit to the spectrum as described in
\S\ref{sec:spec_measure}. For lobe-dominated radio quasars we use all the matched FIRST sources to compute the radio flux density. We note that we
may
have missed some double-lobed radio quasars with lobe diameter larger than 1'.

To flag BALQSOs, we use the \citet{Gibson_etal_2009} DR5 BALQSO catalog to set the \CIV\ and \MgII\ BALQSO flags (using their ``BI0'' flags).  We
also
visually inspected all the post-DR5 quasars with redshift $z>1.45$ to identify obvious \CIV\ BALQSOs (we may have missed some weak BALQSOs). We did
not perform a systematic search for low-ionization \MgII\ BALQSOs because of the large number of quasars with \MgII\ coverage and the much rarer
occurrence of \MgII\ BALQSOs. Although we report serendipitously identified \MgII\ BALQSOs, the completeness of these objects is low. We identified a
total of 6214 BALQSOs in the DR7 quasar catalog.

%\textbf{Add description for disk emitters and double-peaked [OIII] objects}.
There are also subclasses of quasars which show interesting spectral features in their broad or narrow emission line profiles. Some quasars show a
double-peaked or asymmetric broad Balmer line profile, which is commonly interpreted as arising from a relativistic accretion disk around the black
hole \citep[disk emitters, e.g.,][]{Chen_Halpern_1989,Eracleous_Halpern_1994,Strateva_etal_2003} although alternative interpretations exist for some
of these objects, such as a binary SMBH \citep[e.g.,][and references therein]{Gaskell_2009}. Some quasars show double-peaked narrow lines \citep[such
as \OIIIab, e.g.,][]{Liu_etal_2009,Smith_etal_2009,Wang_etal_2009a}, which could be due to either narrow line region kinematics or a merging AGN pair
\citep[e.g.,][]{Liu_etal_2010,Shen_etal_2011a}. We have visually inspected all of the $z<0.89$ quasars in the DR7 catalog (i.e., those with \hbeta\
coverage) and flagged such objects.

\section{Spectral Measurements}\label{sec:spec_measure}

We are primarily interested in the broad \halpha, \hbeta, \MgII, and \CIV\ emission lines because these are the most frequently studied lines
that are available over a wide range of redshifts. More importantly, they have been calibrated as virial black hole (BH) mass estimators
\citep[e.g.,][]{Vestergaard_2002,McLure_Jarvis_2002,McLure_Dunlop_2004,Greene_Ho_2005,
Vestergaard_Peterson_2006,McGill_etal_2008,Vestergaard_Osmer_2009,Wang_etal_2009b}.

There are numerous studies of the statistical emission line
properties of quasars relying either on direct measurements, or on spectral fits of the line profile
\citep[e.g.,][]{Boroson_Green_1992,Marziani_etal_1996,McLure_Jarvis_2002,Richards_etal_2002b,
McLure_Dunlop_2004,Bachev_etal_2004,
Dietrich_Hamann_2004,Baskin_Laor_2005,Kollmeier_etal_2006,
Fine_etal_2006,Fine_etal_2008,Bonning_etal_2007,
Salviander_etal_2007,Sulentic_etal_2007,ShenJ_etal_2008,Shen_etal_2008b,Hu_etal_2008a,
Hu_etal_2008b,Zamfir_etal_2009,Wang_etal_2009b,Wu_etal_2009,Dong_etal_2009a,Dong_etal_2009b}.
For the same set of data, different studies sometimes report
different results for certain measured quantities due to the
different line-measurement techniques used in these studies. Which method
is preferred, however, depends on the nature of the problem under
study. A classic example is measuring the {\em
full-width-at-half-maximum} (FWHM) in estimating the BH mass using
virial estimators, where the usual complications are: 1) how to subtract the continuum underneath the line; 2) how
to treat the narrow line component (especially for \MgII\ and \CIV);
3) how to measure the broad line profile, especially in the presence of noise and absorption. These choices crucially
depend on the particular virial estimator calibrations used; in particular, one must use a similar FWHM definition as was used in the virial
estimator
calibration, and new methods of FWHM measurements must be re-calibrated either against reverberation mapping (RM) masses or internally between
different line estimators. On the other hand, different
line-measurement methods have different sensitivities to the quality of
the spectra (spectral resolution and S/N), which
introduce systematics when switching from high-quality to
low-quality data \citep[e.g.,][]{Denney_etal_2009}. It is beyond
our scope to fully settle these issues within the current study.

We remove the effects of Galactic extinction in the SDSS spectra using the \citet{SFD_1998} map and a Milky Way extinction curve from
\citet{CCM_1989}
with $R_V=3.1$, and shift the spectra to restframe using the cataloged redshift as the systemic redshift\footnote{\citet{Hewett_Wild_2010} provided
improved redshifts for SDSS quasars. However, this subtlety has negligible effects on our spectral fits.}. For each line, we fit a local power-law
continuum ($f_\lambda=A\lambda^{\alpha_\lambda}$) plus an iron template
\citep[][]{Boroson_Green_1992,Vestergaard_Wilkes_2001,Salviander_etal_2007}
to the wavelength range around the line that is not contaminated by the broad line emission. During
the continuum$+$iron fitting we simultaneously fit five
parameters: the normalization $A$ and slope $\alpha_\lambda$ of
the power-law continuum, and the normalization $A_{\rm Fe}$, line
broadening $\sigma_{\rm Fe}$ and velocity offset $v_{\rm Fe}$
relative to the systemic redshift for the iron template fit.
Because of the moderate spectral quality of SDSS spectra (median
S/N$\la 10$) $\sigma_{\rm Fe}$ and $v_{\rm Fe}$ are often poorly constrained; nevertheless the iron fit
gives a reasonably good estimate of the iron flux to be subtracted
off. The continuum$+$iron fit is then subtracted from the
spectrum, and the resulting line spectrum is modelled by various
functions. In the case of \halpha\ and \hbeta\ the adjacent narrow emission
lines, e.g., \OIIIab, \NIIab, \SIIab, are also fit simultaneously.
Below we describe the detailed fitting procedures for the four
broad lines.

\subsection{\halpha}\label{subsec:ha}
For \halpha\ we use the optical iron template from
\citet{Boroson_Green_1992}, and we fit for objects with $z\le
0.39$. The continuum$+$iron fitting windows are [6000,6250] \AA\
and [6800,7000] \AA.

For \halpha\ line fitting, we fit the wavelength range [6400,6800]
\AA. The narrow components of \halpha, \NIIab, \SIIab\ are each fit
with a single Gaussian. Their velocity offsets from the systemic
redshift and line widths are constrained to be the same, and the relative
flux ratio of the two \NII\ components is fixed to $2.96$. We impose an
upper limit on the narrow line FWHM $< 1200\ {\rm
km\,s^{-1}}$ \citep[e.g.,][]{Hao_etal_2005}. The broad \halpha\ component is modelled in two different ways: a) a single Gaussian with a FWHM $>1200\
{\rm km\,s^{-1}}$; b) multiple Gaussians with up to three Gaussians, each with a FWHM $>1200\ {\rm km\,s^{-1}}$.
The second method yields similar results to the fits with a
truncated Gaussian-Hermite function
\citep[e.g.,][]{Vandermarel_etal_1993}. During the fitting, all
lines are restricted to be emission lines (i.e., positive flux).
%We used initial guesses of $\sigma=3000 \ {\rm km\,s^{-1}}$ for the broad components and $\sigma=300\ {\rm km\,s^{-1}}$ for
%the narrow components. We used the vacuum wavelengths of the emission lines as initial guesses for line centroids, and arbitrary initial guesses for
%the line fluxes. Fitting results are generally insensitive to the initial values used in the fitting.
%We measure line properties such as velocity offset with respect to the systemic redshift,
%FWHM, equivalent width (EW) from the best fits of the line
%profile.

%One problem with the \hbeta+\OIIIab\ fitting procedure is the
%usage of a single Gaussian for the \OIIIab\ line profiles. Since
%\OIIIab\ frequently show asymmetric blue wings
%\citep[e.g.,][]{Heckman_etal_1981,Greene_Ho_2005a,Komossa_etal_2008b}
%and sometimes even more dramatic double-peaked profiles
%\citep[e.g.,][]{Liu_etal_2009,Smith_etal_2009,Wang_etal_2009a}, a
%single Gaussian fit will degrade the overall fitting quality for
%\OIIIab, and lead to problematic subtraction of the narrow \hbeta\
%flux and bias the broad \hbeta\ measurement. Nevertheless the
%fraction of such incidents is low ({\bf $\la 1\%$}). In cases
%where the narrow \OIIIab, or \NIIab, \SIIab\ are weak, the fits
%could also lead to problematic narrow \halpha/\hbeta\ subtraction
%-- {\bf these cases constitute $\la 1\%$ of the sample.}

\subsection{\hbeta}\label{subsec:hb}
For \hbeta\ we use the optical iron template from
\citet{Boroson_Green_1992}, and we fit for objects with $z\le
0.89$. The continuum$+$iron fitting windows are [4435,4700] \AA\
and [5100,5535] \AA. For the \hbeta\ line fitting, we follow a
similar procedure as \halpha\ to fit for \hbeta\ and \OIIIab,
where the line fitting wavelength range is [4700,5100] \AA. Since the
\OIIIab\ lines frequently show asymmetric blue wings
\citep[e.g.,][]{Heckman_etal_1981,Greene_Ho_2005a,Komossa_etal_2008b}
and sometimes even more dramatic double-peaked profiles
\citep[e.g.,][]{Liu_etal_2009,Smith_etal_2009,Wang_etal_2009a}, we
model each of the narrow \OIIIab\ lines with two Gaussians, one for
the core and the other for the blue wing. The flux ratio of the \OIII\ doublet is not fixed during the fit, but we found that the fitting results
show
good
agreement with the theoretical ratio of about 3. The velocity offset and
FWHM of the narrow \hbeta\ line are tied to those of the core \OIIIab\
components\footnote{In some objects the narrow \hbeta\ component might have different width and
velocity offset from those of the core \OIIIab\ components. Usually in such cases our procedure still
provides a reasonable approximation to subtract the narrow \hbeta\ component; however, under rare circumstances
where the narrow \hbeta\ component is not obvious this may lead to a biased narrow line subtraction. Nevertheless, fitting the narrow \hbeta\ line
without this constraint would significantly degrade the reliability of narrow line subtraction. }, and we impose an upper limit of $1200\ {\rm
km\,s^{-1}}$ on the narrow line FWHM. As in the \halpha\ case, the
broad \hbeta\ component is modelled in two ways: either by a single Gaussian, or by multiple Gaussians with up to three Gaussians, each with a FWHM
$>1200 \ {\rm
km\,s^{-1}}$.

The single Gaussian fit to the broad component is
essentially the same to the procedure in \citet{Shen_etal_2008b}, and is
somewhat similar to the procedure in
\citet{McLure_Dunlop_2004}\footnote{In addition to Gaussian profiles,
\citet{McLure_Dunlop_2004} also tried to fit the broad/narrow
component with a single Lorentzian, but this does not
change the measured broad FWHM significantly.}. However, in many objects
the broad \halpha/\hbeta\ component cannot be fit perfectly with a
single Gaussian; and FWHMs from the single Gaussian fits are
systematically larger by $\sim 0.1$ dex than those from the multiple Gaussian fits
\citep[e.g.,][]{Shen_etal_2008b}. The additional
multiple Gaussian fits for the broad \halpha/\hbeta\ component
provide a better fit to the
overall broad line profile, and the FWHM measured from the model
flux can be used in customized virial calibrations. It is unclear,
however, which FWHM is a better surrogate for the true virial velocity,
that is, the one that yields the smallest scatter in the calibration
against RM black hole masses.

%{\bf Show several comparison between different FWHMs, i.e., single
%Gaussian against multiple Gaussian, Greene \& Ho for \halpha.}

\subsection{\MgII}\label{subsec:mgii}
For \MgII\ we use the UV iron template from
\citet[][]{Vestergaard_Wilkes_2001}%\citep[and the updated version in][]{Salviander_etal_2007}
, and we fit for objects with $0.35\le
z\le 2.25$. The continuum$+$iron fitting windows are [2200,2700]
\AA\ and [2900,3090] \AA. We then subtract the pseudo-continuum
from the spectrum, and fit for the \MgII\
line over the [2700,2900] \AA\ wavelength range, with a single Gaussian
(with ${\rm FWHM}<1200\ {\rm kms^{-1}}$) for the narrow \MgII\
component, and for the broad \MgII\ component with: 1) a single
Gaussian; 2) multiple Gaussians with up to three Gaussians. Again, the multiple-Gaussian fits often provide a better fit to the overall broad \MgII\
profile; but we have retained the FWHMs from a single Gaussian fit
in order to use the \MgII\ virial mass calibrations in
\citet[][]{McLure_Jarvis_2002} and \citet{McLure_Dunlop_2004}.
Some \MgII\ virial estimator calibrations
\citep[e.g.,][]{McLure_Jarvis_2002,McLure_Dunlop_2004,Wang_etal_2009b}
do subtract a narrow \MgII\ component while others
\citep[e.g.,][]{Vestergaard_Osmer_2009} do not. To utilize the
\MgII\ calibration in \citet{Vestergaard_Osmer_2009}, we also
measure the FWHMs from the broad+narrow \MgII\ fits (with multiple
Gaussians for the broad component), where any Gaussian component
having flux less than $5\%$ of the total line flux is rejected
when computing the FWHM --- this step is to eliminate artificial
noise spikes which can bias the FWHM measurements. During our fitting, we
mask out $3\sigma$ outliers below the 20-pixel boxcar-smoothed spectrum to reduce
the effects of narrow absorption troughs.

\begin{figure}
  \centering
    \includegraphics[width=0.45\textwidth]{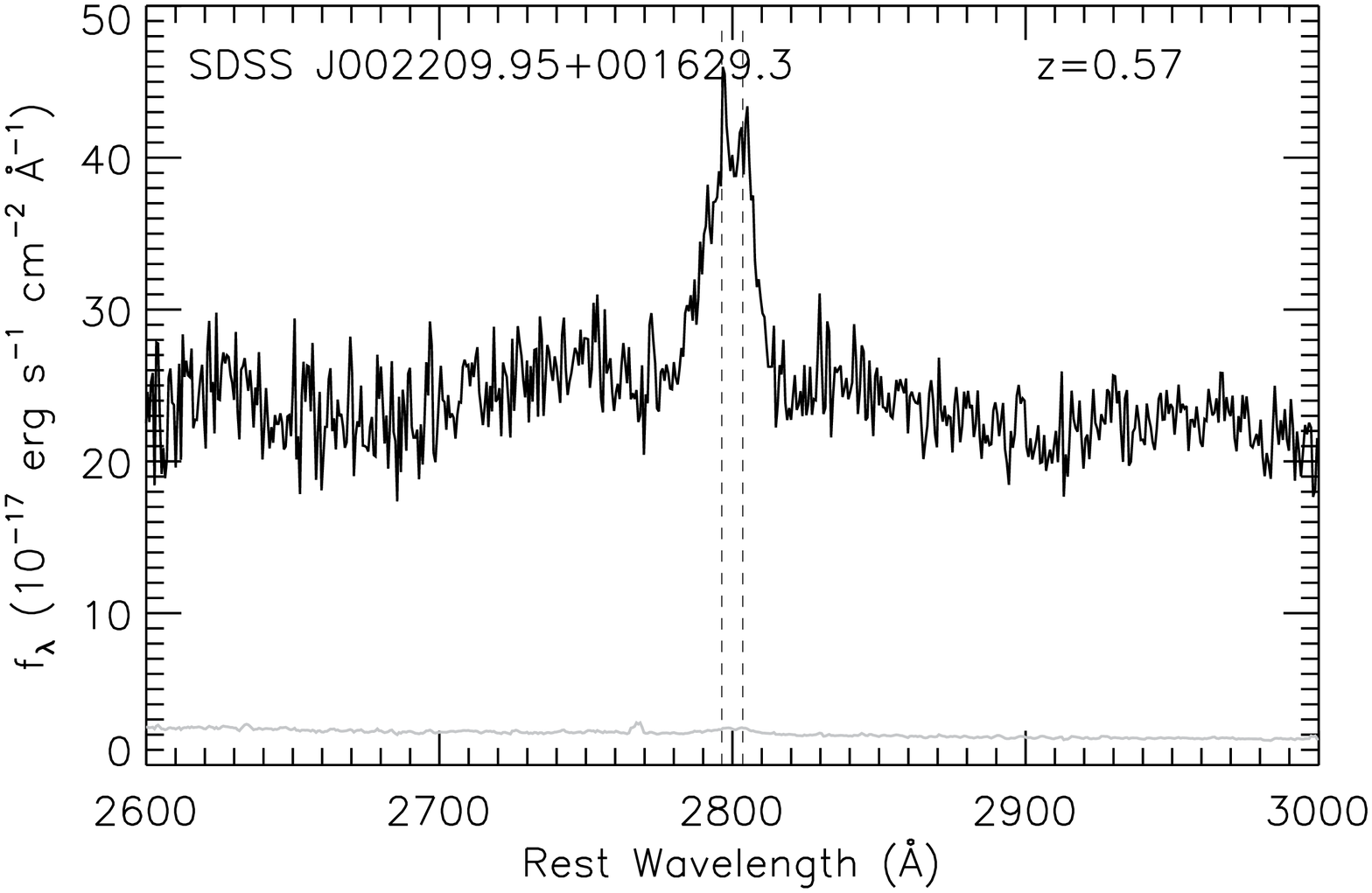}
    \includegraphics[width=0.45\textwidth]{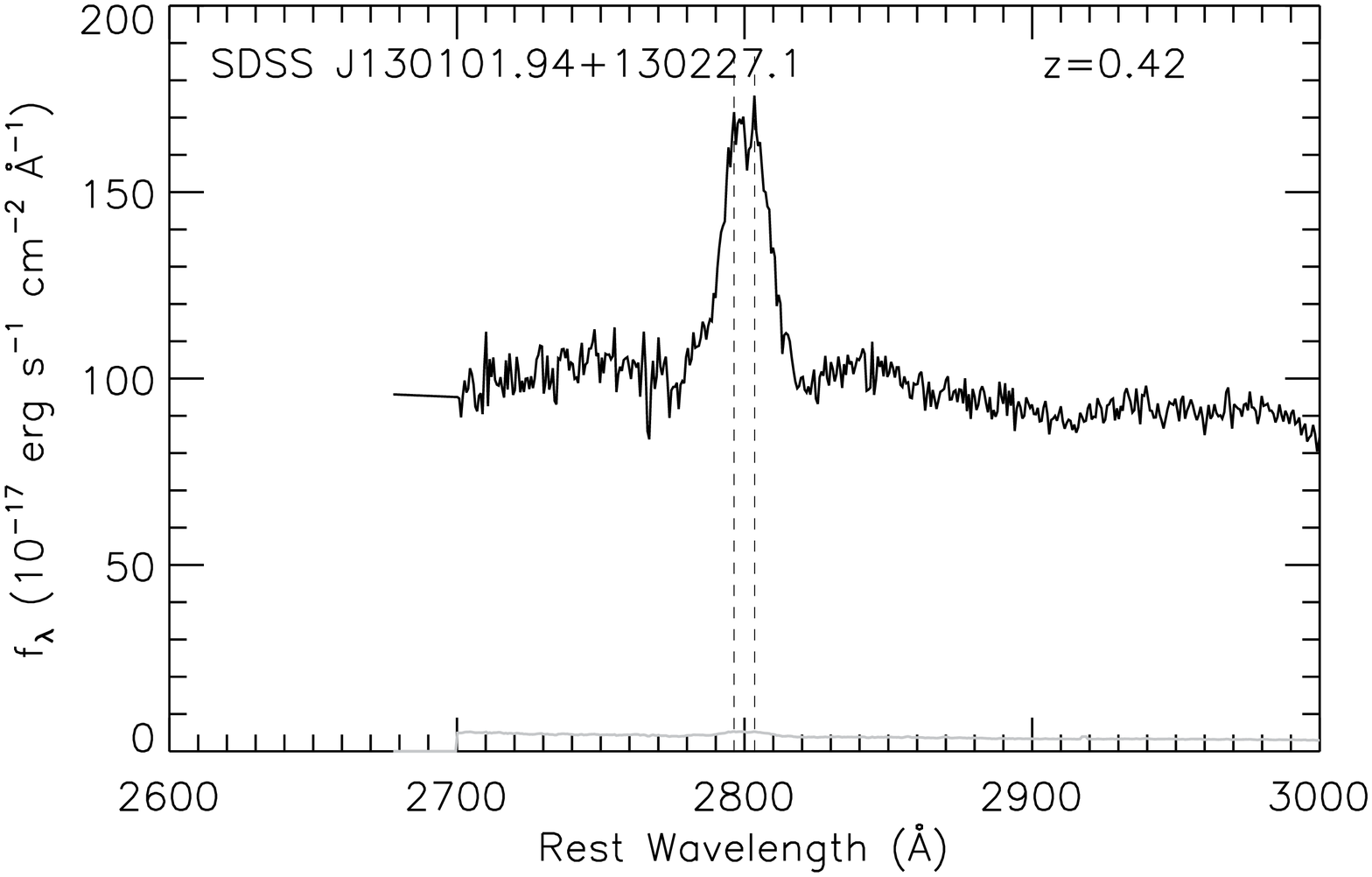}
    \caption{Two examples of \MgII\ lines which show narrow line components. The spectra are plotted as black lines and the gray lines show the
    errors
    per pixel. The dashed vertical lines mark the locations of the \MgII\ $\lambda\lambda$2796,2803 doublet.}
    \label{fig:mgii_examp}
\end{figure}

\begin{figure*}
  \centering
    \includegraphics[width=1\textwidth]{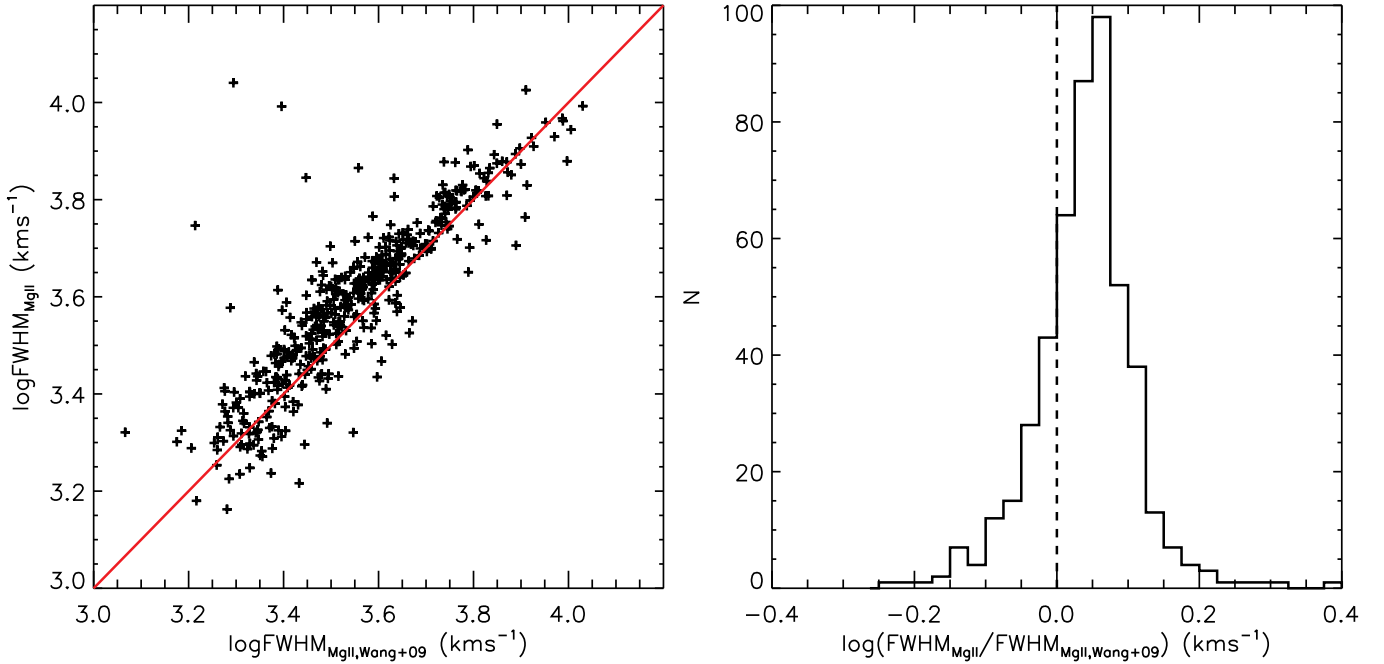}
    \caption{Comparison of our \MgII\ FWHMs with those from \citet{Wang_etal_2009b} for the same objects. The left panel shows a scatter plot, where
    the solid line is the
    unity relation. The right panel shows a histogram of the ratio between the two values. Our broad \MgII\ FWHM values are systematically larger by
    $\sim 0.05$ dex than those in \citet{Wang_etal_2009b},
    mainly caused by the fact that they fit the broad \MgII\ as a doublet while we did not.}
    \label{fig:comp_mgii_wang09}
\end{figure*}

Unlike the cases of \halpha\ and \hbeta, it is somewhat
ambiguous whether it is necessary to subtract a narrow
line component for \MgII\ and if so, how to do it. On one hand, for some objects, such as
SDSSJ002209.95+001629.3 and SDSSJ130101.94+130227.1 (e.g., Fig.\
\ref{fig:mgii_examp}), the spectral quality is sufficient to see
the bifurcation of the \MgII\ doublet around the peak. The locations
of the two peaks indicate that they are associated with the \MgII\
$\lambda\lambda$2796,2803 doublet, and the fact that they are
resolved means that the FWHM of each component is $\la 750\ {\rm
km\,s^{-1}}$, hence they are most likely associated with the narrow line region. On the other hand, such cases are rare and most SDSS spectra
do not have adequate S/N to unambiguously locate the narrow
\MgII\ doublet. Associated narrow \MgII\ absorption troughs can further complicate the situation by mimicing two peaks. Hence although our approach
of
fitting a single Gaussian to the narrow \MgII\ component is not perfect, it nevertheless accounts for some narrow \MgII\ contamination. Fig.\
\ref{fig:comp_mgii_wang09} compares our broad
\MgII\ FWHM measurements with those from \citet{Wang_etal_2009b} for the objects in both studies. Although we have used a different approach, our
results are consistent with theirs, with a mean offset $\sim 0.05$ dex. This systematic offset between our results and theirs is caused by the fact
that they are treating the broad \MgII\ line as a doublet as well, while we (and most studies) are treating the broad \MgII\ line as a single
component.

\begin{figure}
  \centering
    \includegraphics[width=0.45\textwidth]{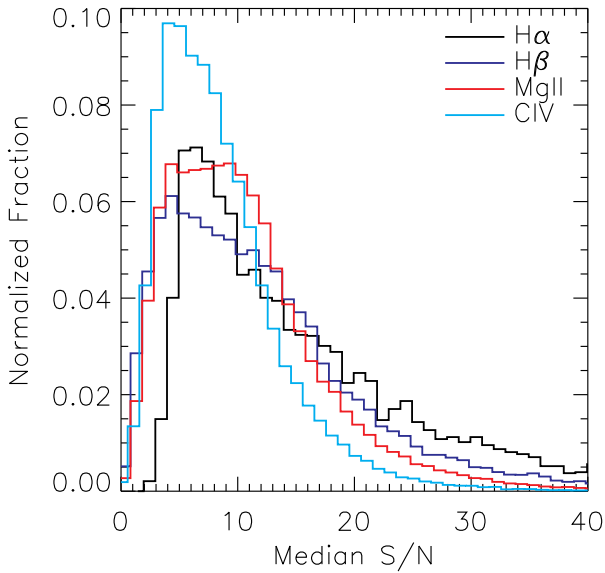}
    \caption{Distributions of the median S/N per pixel around the line-fitting regions for objects with line measurements, for \halpha, \hbeta,
     \MgII\ and \CIV\ respectively, where the underlying area of each distribution is normalized at unity. }
    \label{fig:SN_dist}
\end{figure}

%In fact, there is no conclusive evidence for a strong narrow line
%\MgII\ component (although in some cases there do appear to have a
%narrow \MgII\ component, {\bf showing an example? Such as
%SDSSJ131459.75+505932.9}).

\subsection{\CIV}\label{subsec:civ}

%For \CIV\ we again use the \citet[][]{Vestergaard_Wilkes_2001}
%iron template, and we fit for objects with $1.45\le z\le 4.95$.

% Note that I fitted objects with $1.45\le z\le 4.95$, but only
% report objects with $1.5\le z\le 4.95$ for CIV, to reduce junk fits at
% the edge of the spectrum

For \CIV\ we fit for objects with $1.5\le z\le 4.95$. Iron emission is generally weak for \CIV, and most of our objects do not have the spectral
quality sufficient for a reliable iron template subtraction
\citep[e.g.,][]{Shen_etal_2008b}. The power-law continuum fitting
windows are: [1445,1465] \AA\ and [1700,1705] \AA. We found
fitting \CIV\ with iron subtraction does not change the fitted
\CIV\ FWHM significantly, but does systematically reduce the \CIV\ EW by $\sim 0.05$ dex because the iron flux under the wings of the \CIV\ line is
accounted for. At the same time, fitting iron emission increases
the uncertainty in the fitted continuum slope and normalization, due to
imperfect subtraction of the iron flux. Therefore
we report our \CIV\ measurements without the iron template fits,
and emphasize that the \CIV\ EWs may be overestimated by $\sim
0.05$ dex on average.

The continuum subtracted line emission within [1500,1600] \AA\ was
fitted with three Gaussians
\citep[e.g.,][]{Shen_etal_2008b}, and we measure the line FWHM
from the model fit. To reduce the effects of noise spikes, we
reject any Gaussian component having flux less than $5\%$ of the
total model flux when computing the FWHM. However, unlike some
attempts in the literature
\citep[e.g.,][]{Bachev_etal_2004,Baskin_Laor_2005,Sulentic_etal_2007,Zamfir_etal_2009},
we do not subtract a narrow \CIV\ component because: 1) it is still debatable if a strong narrow \CIV\ component exists for most quasars, or if it is
feasible to do such a subtraction; 2) existing \CIV\ virial estimators are calibrated using the FWHMs from the entire \CIV\ profile
\citep[][]{Vestergaard_Peterson_2006}.

Many \CIV\ lines are affected by narrow or broad absorption features.
To reduce the effects of such absorption on the \CIV\ fits, we mask out
$3\sigma$ outliers below the 20-pixel boxcar-smoothed spectrum
during our fits (to remedy for narrow absorption features); we also perform a second fit excluding pixels
below $3\sigma$ of the first model fit, and replace the first one
if statistically justified (to account for broad absorption features). We found these recipes can alleviate the impact of narrow or moderate
absorption features, but the improvement is marginal for objects severely affected by broad absorption.

%{\bf Describe caveats with \CIV, such as poor spectral quality,
%inherent complex line profile, etc.}

\subsection{Reliability of spectral fits and error estimation}\label{subsec:fits_quality}

Our spectral fits were performed in an automatic fashion. Upon
visual inspection of the fitting results we are confident that
the vast majority ($\ga 95\%$) of the fits to high S/N spectra
were successful, and comparisons with independent fits by
others also show good agreement. However, the reliability of
our spectral fits drops rapidly for low-quality spectra. Fig.\
\ref{fig:SN_dist} shows the distributions of the median S/N per
pixel around the line-fitting region for objects that have line
measurements, for \halpha, \hbeta, \MgII\ and \CIV\
respectively. Although the bulk of objects have median S/N$>5$
for the line-fitting regions, there are many objects that have
lower median S/N, especially for \CIV\ at high redshift. The
effects of S/N on the measurements depend on both the
properties of the lines (i.e., line profile, line strength,
degree of absorption features, etc), and the line-fitting
technique itself (i.e., what functional form was used, how to
deal with absorption troughs, etc).

To investigate the impact of S/N on our fitting parameters we ran a series of Monte Carlo simulations. We select representative real spectra with
high
S/N, then degrade the spectra by adding Gaussian noise and measure the line properties using the same line-fitting routine. For each line (\hbeta,
\MgII, or \CIV), we study several objects with various line shapes and EWs. We simulate 500 trials for each S/N level and take the median and the
$68\%$ range as the measurement result and its error.

\begin{figure}
  \centering
    \includegraphics[width=0.48\textwidth]{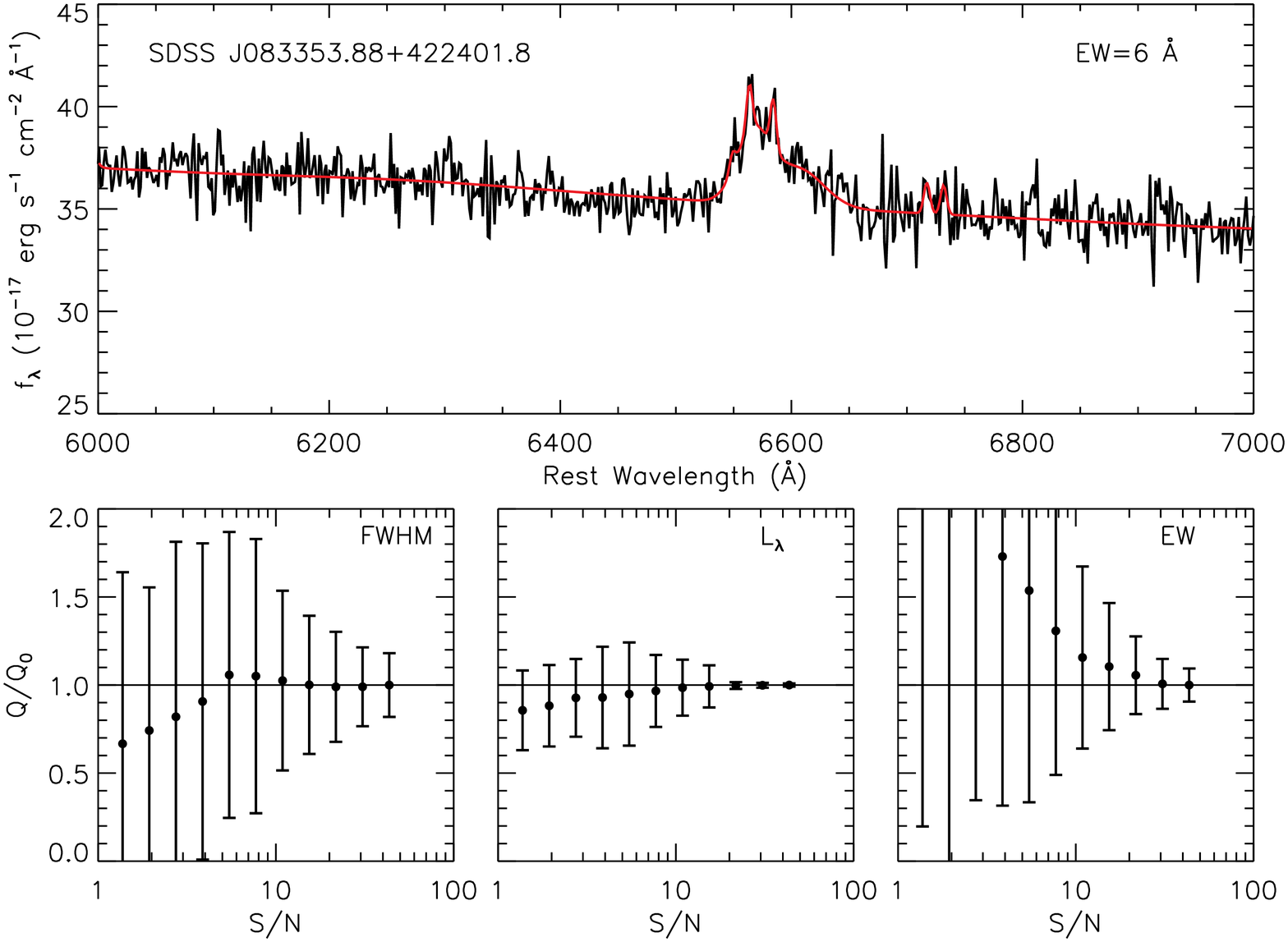}
    \includegraphics[width=0.48\textwidth]{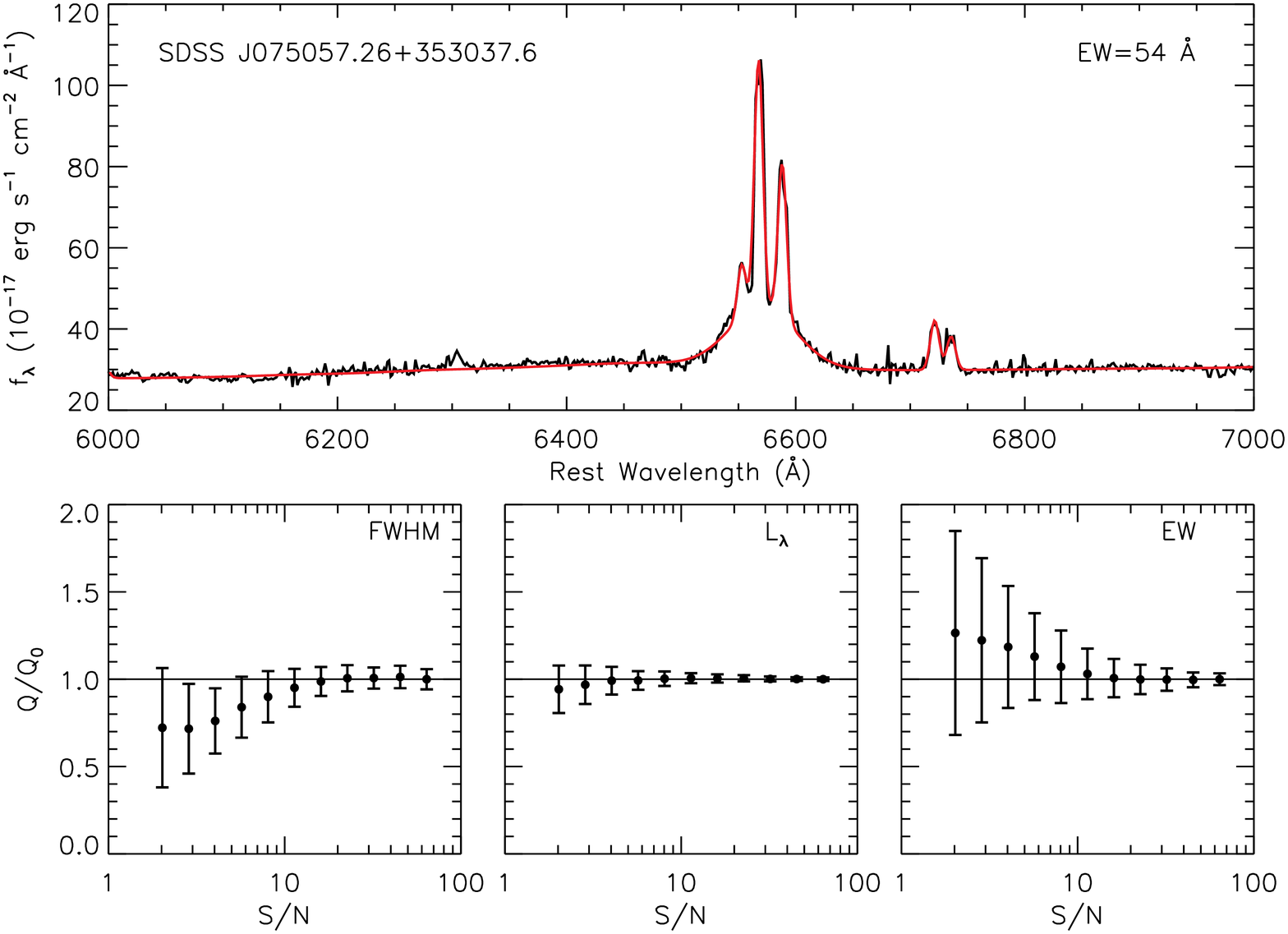}
    \includegraphics[width=0.48\textwidth]{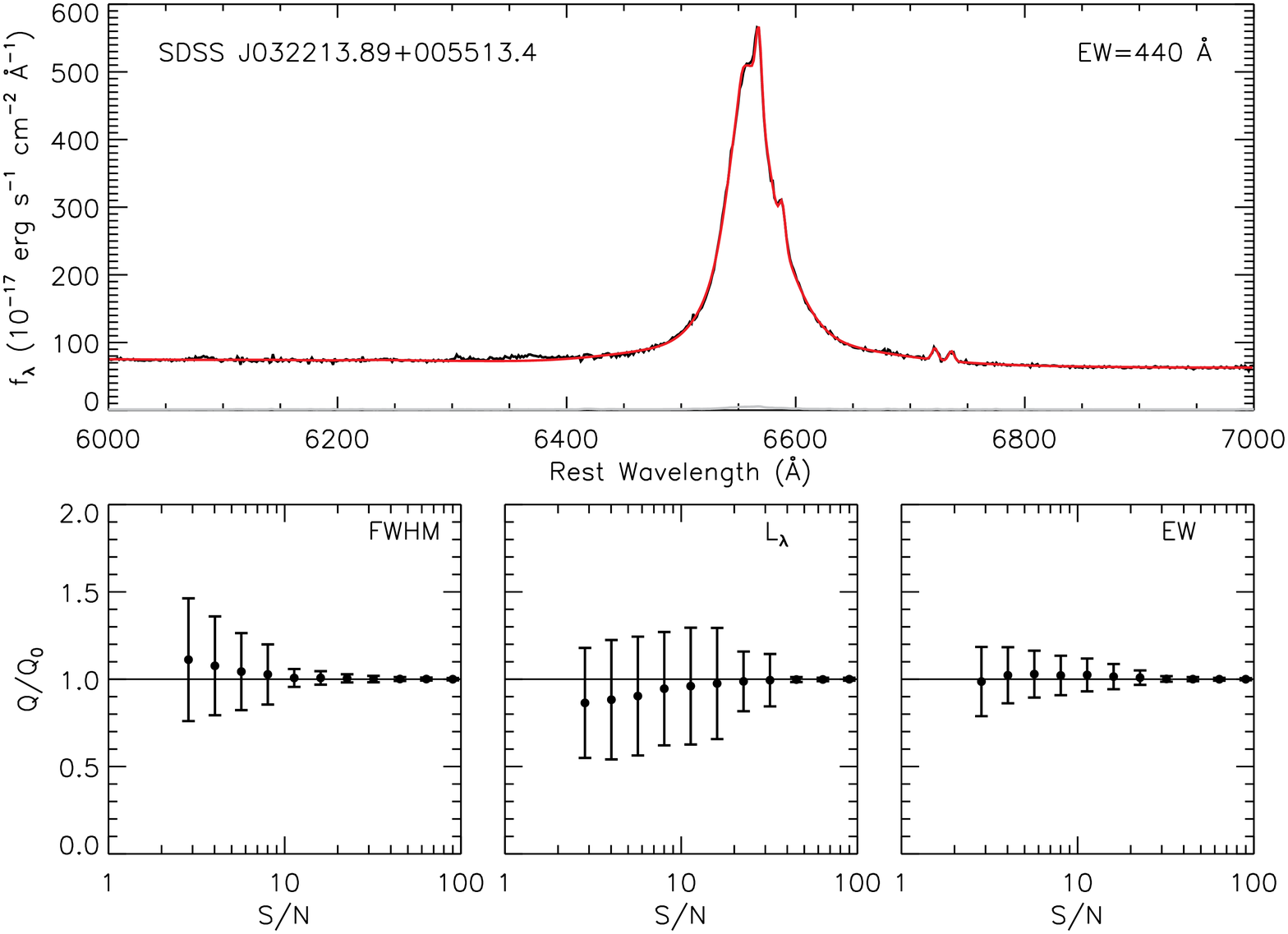}
    \caption{Effects of S/N on the line measurements for \halpha\ for three representative examples. For each object we show
    the actual spectrum (black line) and the best-fit model (red line) in the upper panel. The lower
    three panels show the ratios of the values measured from the degraded spectra to those measured
    from the original spectrum, as functions of S/N; black dots are median values and the error bars indicate the 68\%
    quantile.}
    \label{fig:SN_effect_halpha}
\end{figure}

\begin{figure}
  \centering
    \includegraphics[width=0.48\textwidth]{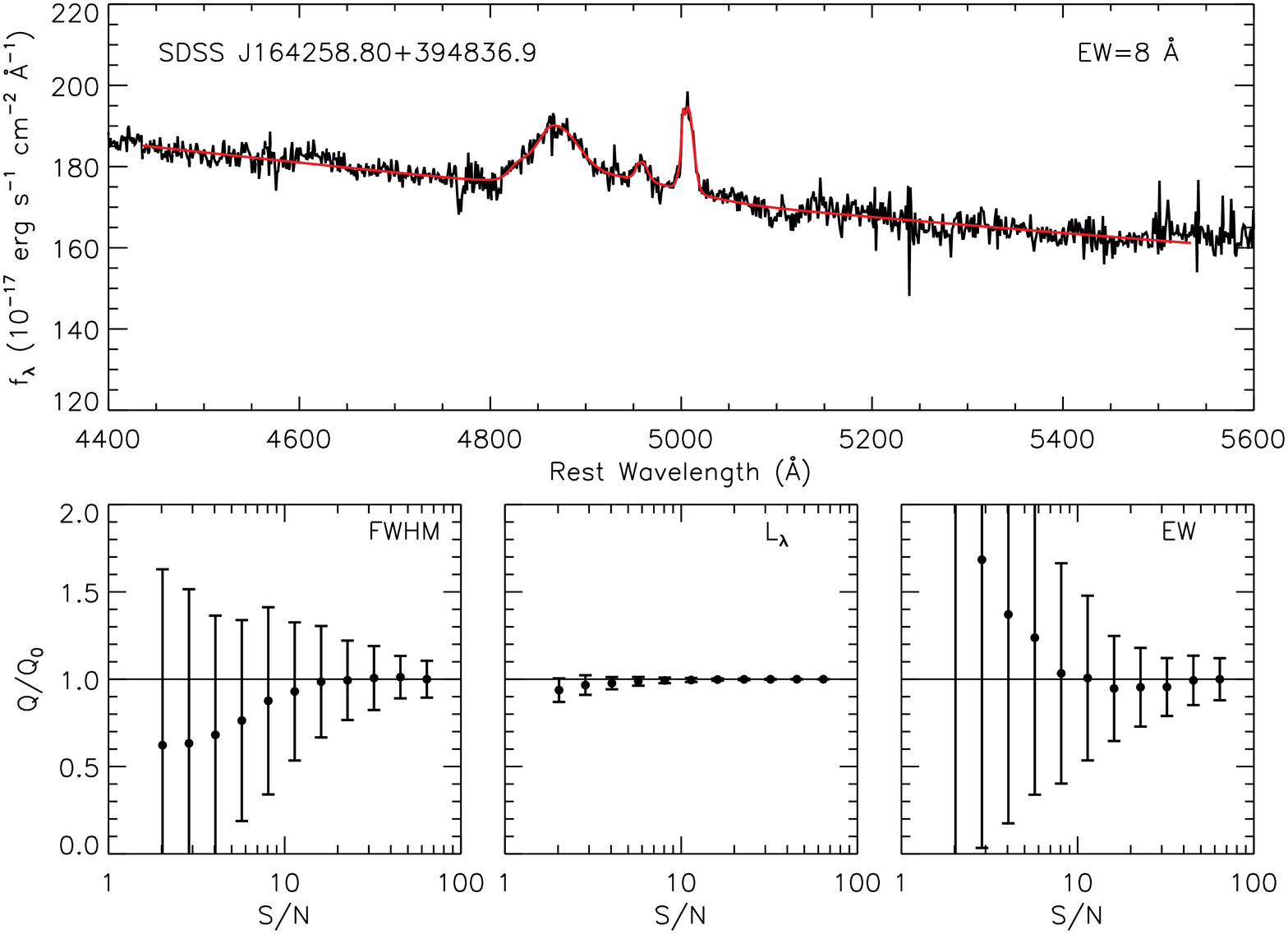}
    \includegraphics[width=0.48\textwidth]{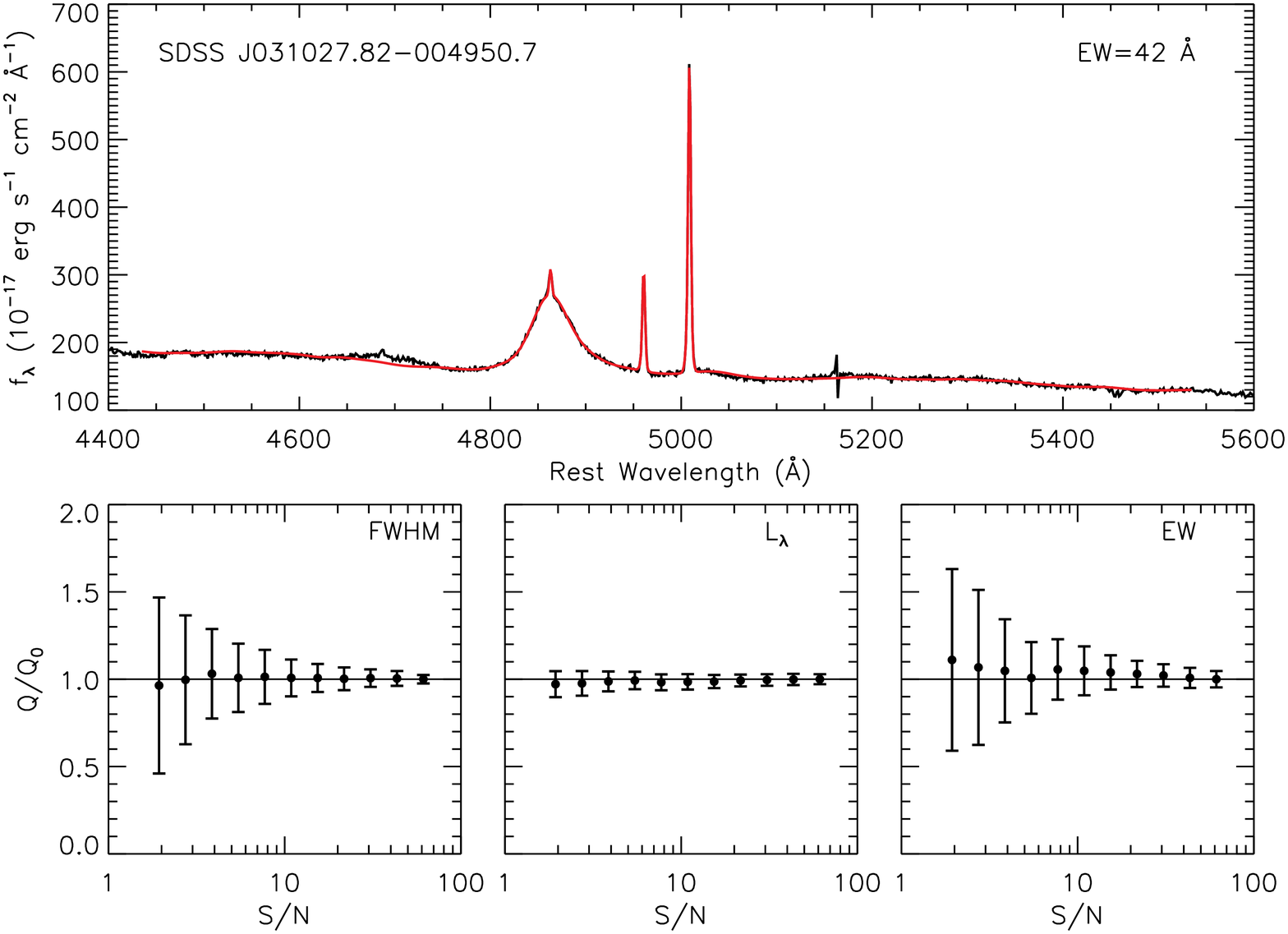}
    \includegraphics[width=0.48\textwidth]{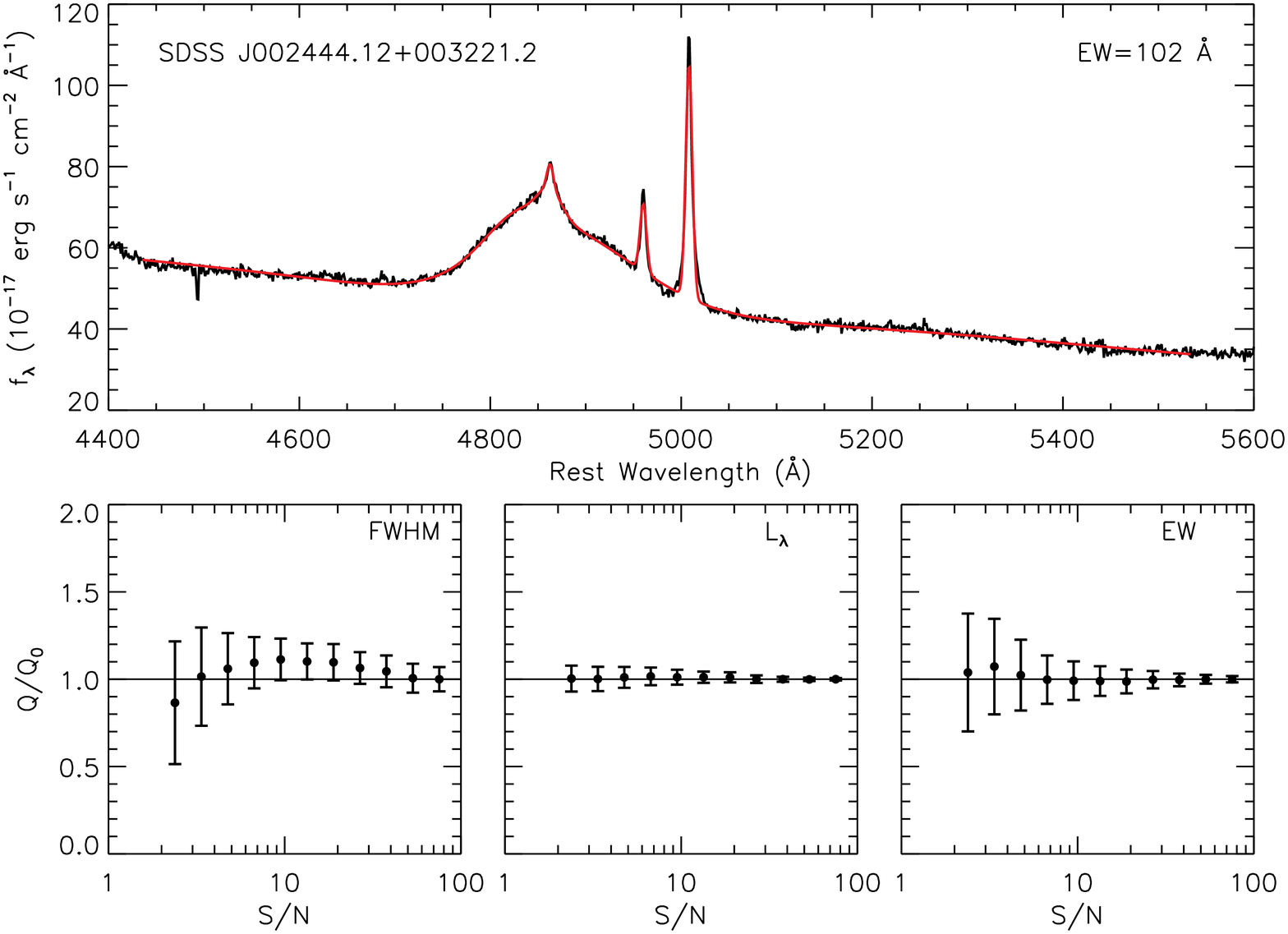}
    \caption{Effects of S/N on the line measurements for \hbeta\ for three representative examples. For each object we show
    the actual spectrum (black line) and the best-fit model (red line) in the upper panel. The lower
    three panels show the ratios of the values measured from the degraded spectra to those measured
    from the original spectrum, as functions of S/N; black dots are median values and the error bars indicate the 68\%
    quantile.}
    \label{fig:SN_effect_hbeta}
\end{figure}

\begin{figure}
  \centering
    \includegraphics[width=0.48\textwidth]{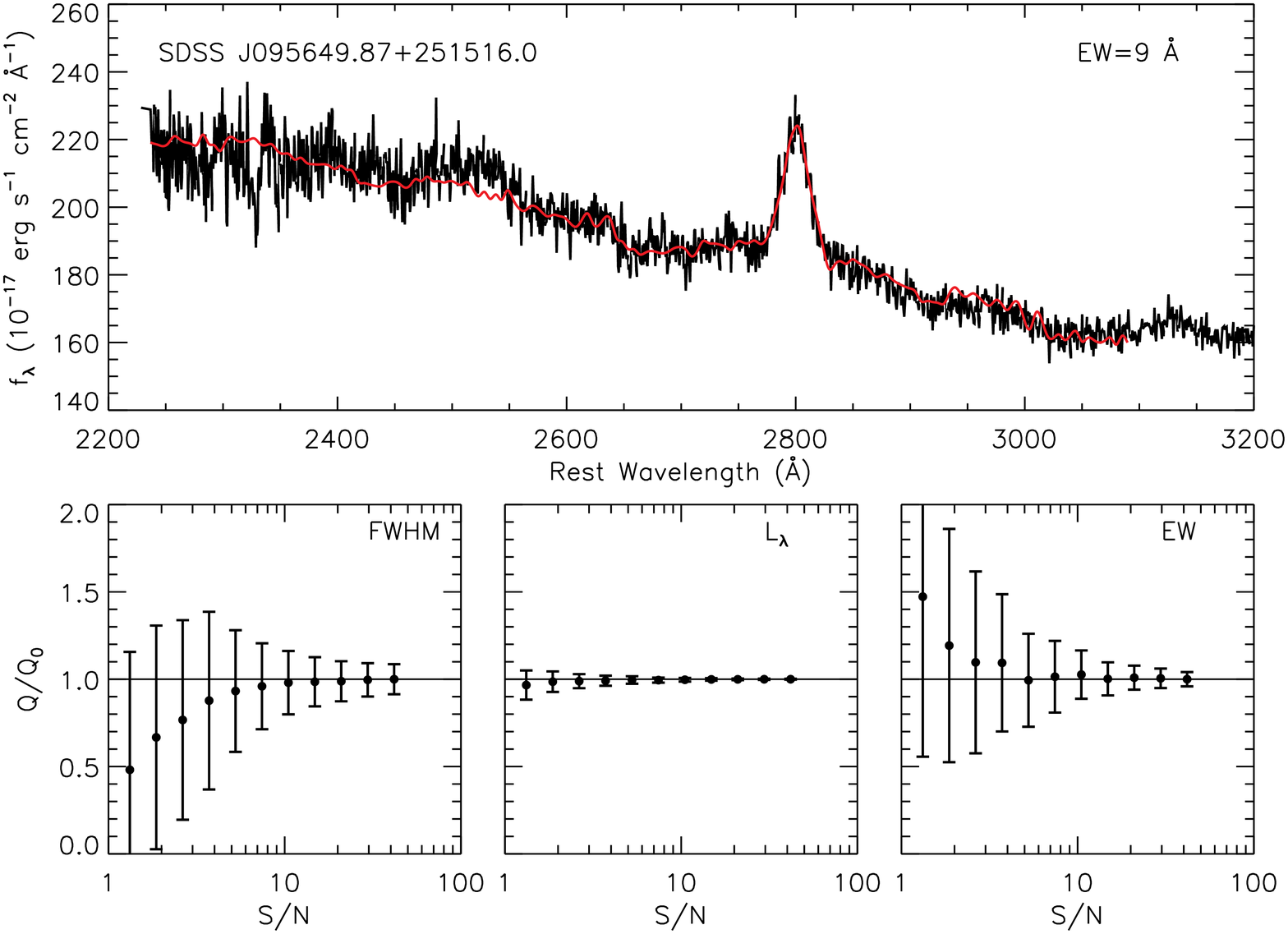}
    \includegraphics[width=0.48\textwidth]{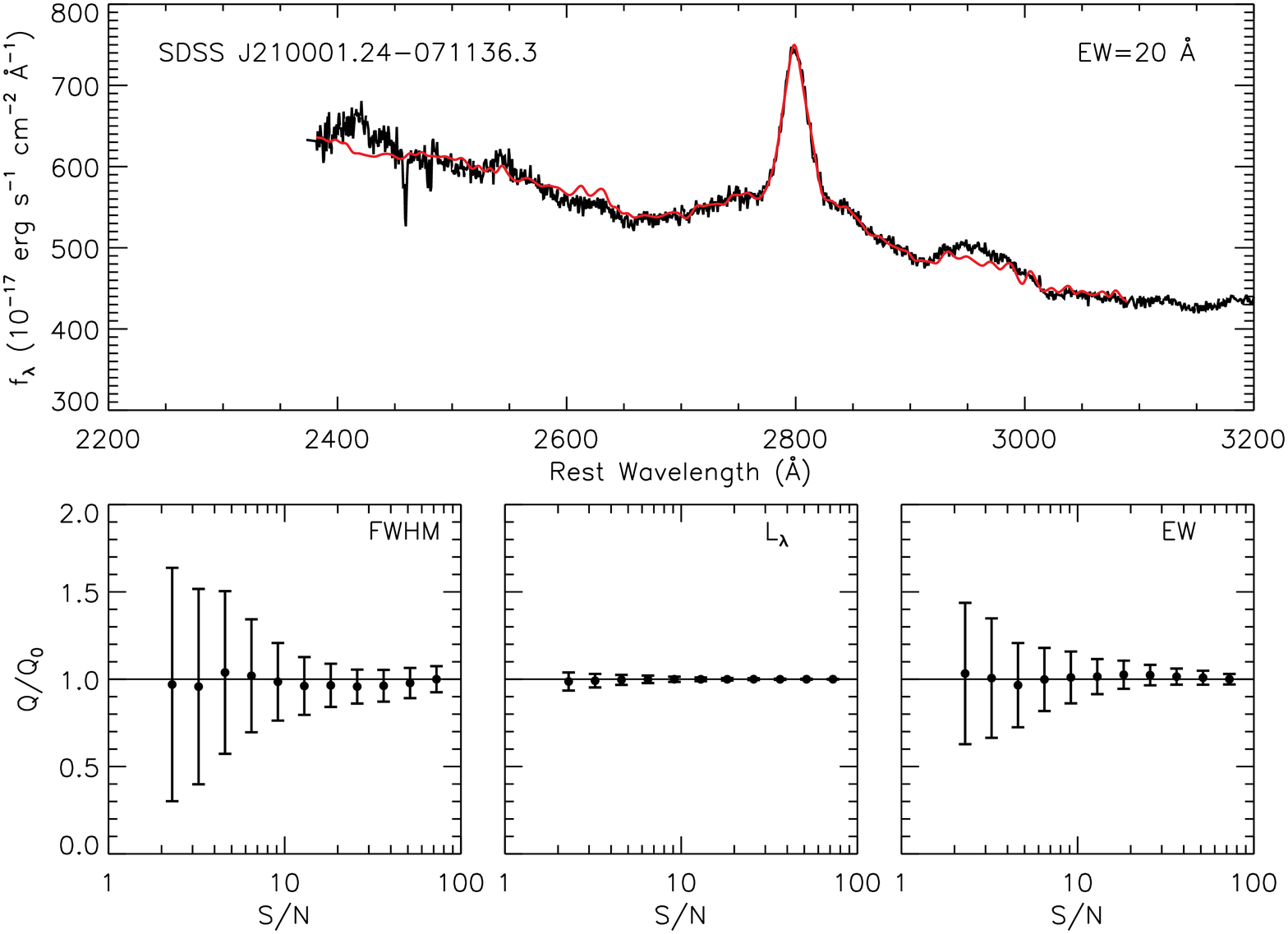}
    \includegraphics[width=0.48\textwidth]{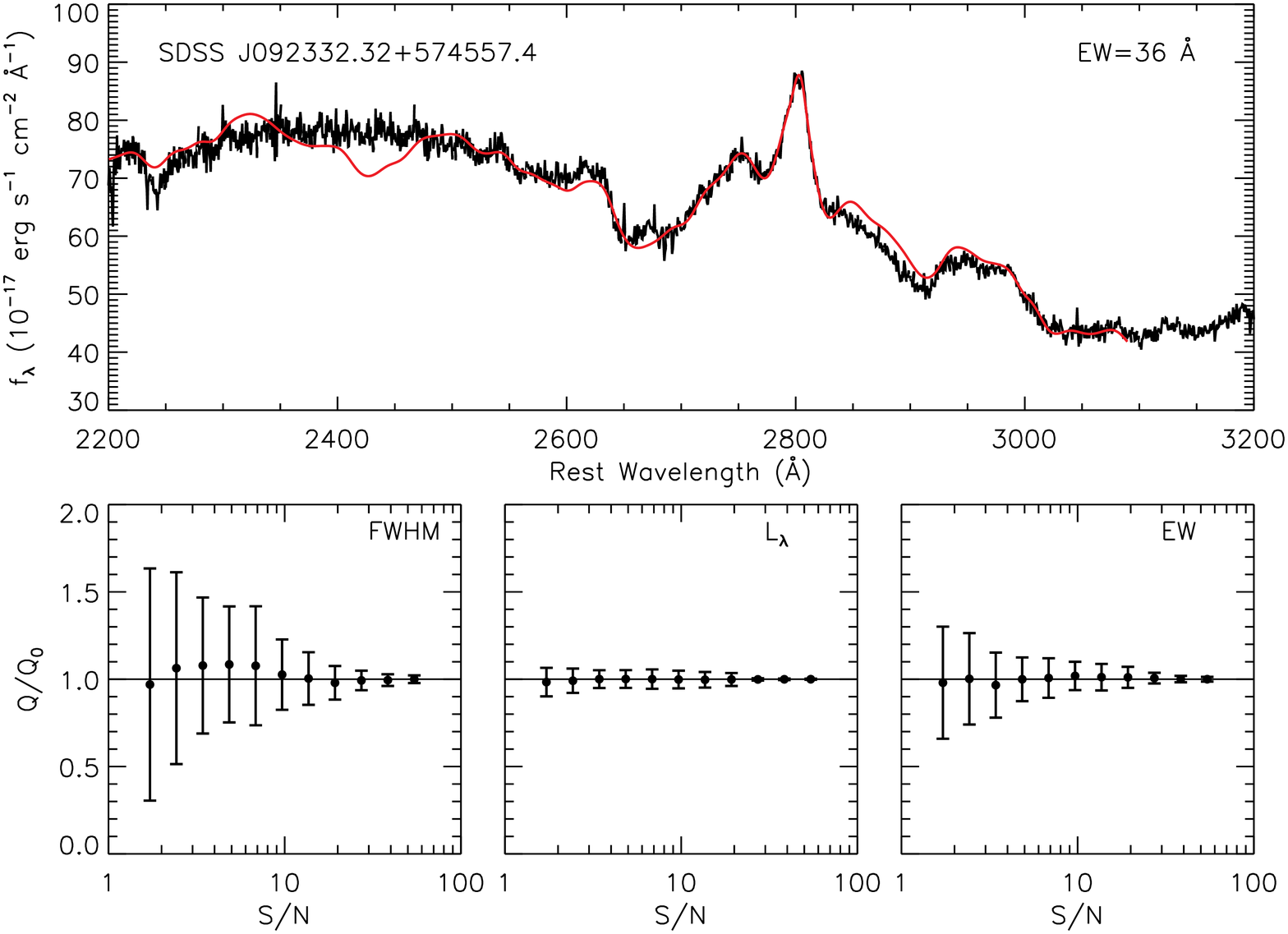}
    \caption{Effects of S/N on the line measurements for \MgII\ for three representative examples. For each object we show
    the actual spectrum (black line) and the best-fit model (red line) in the upper panel. The lower
    three panels show the ratios of the values measured from the degraded spectra to those measured
    from the original spectrum, as functions of S/N; black dots are median values and the error bars indicate the 68\%
    quantile. Note that our continuum$+$iron fit does not account for the [Ne\,{\tiny IV}]/Fe\,{\tiny III} and [O\,{\tiny II}] emission around
    2400\,\AA-2480\,\AA. }
    \label{fig:SN_effect_mgii}
\end{figure}

\begin{figure}
  \centering
    \includegraphics[width=0.48\textwidth]{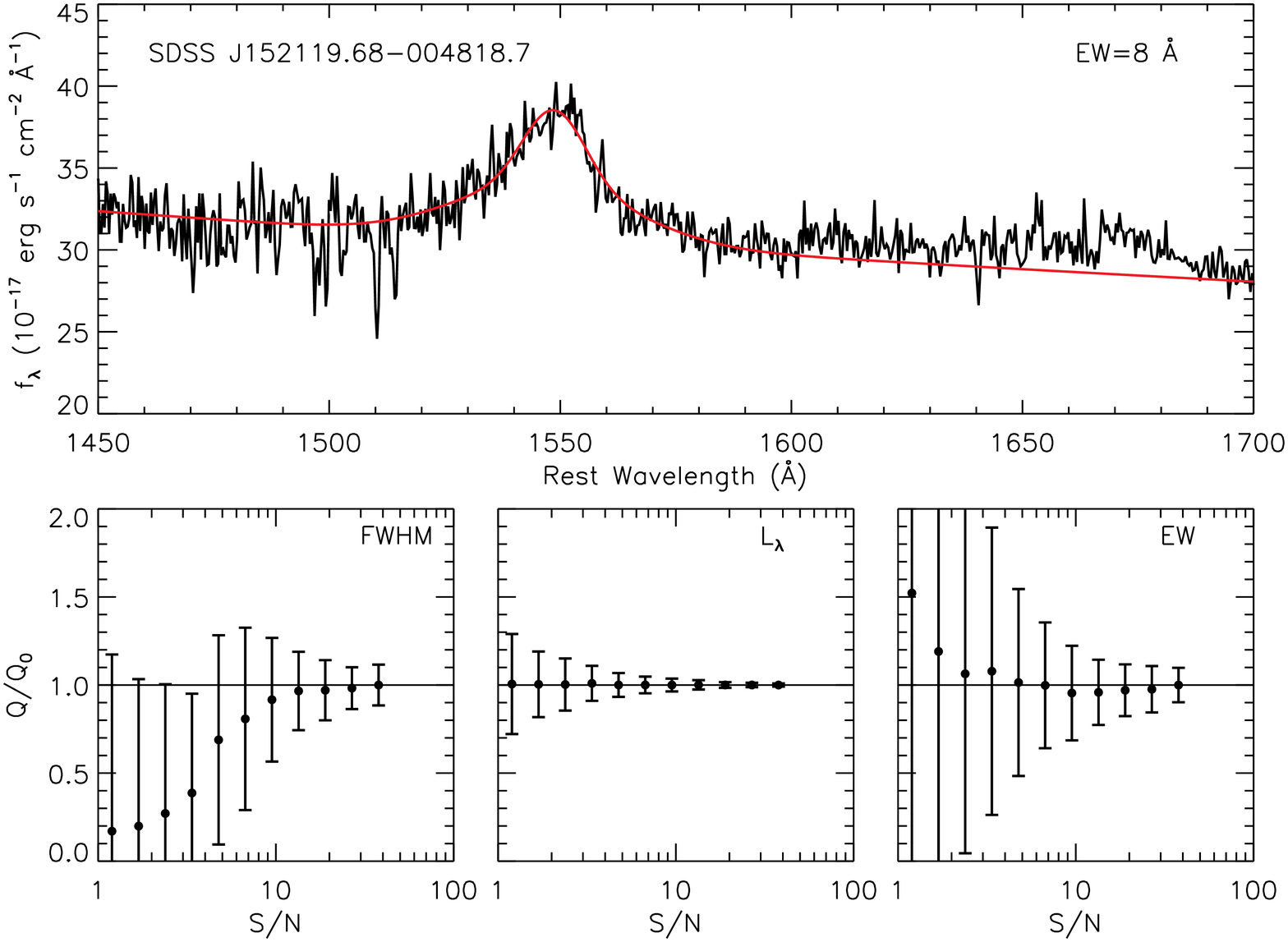}
    \includegraphics[width=0.48\textwidth]{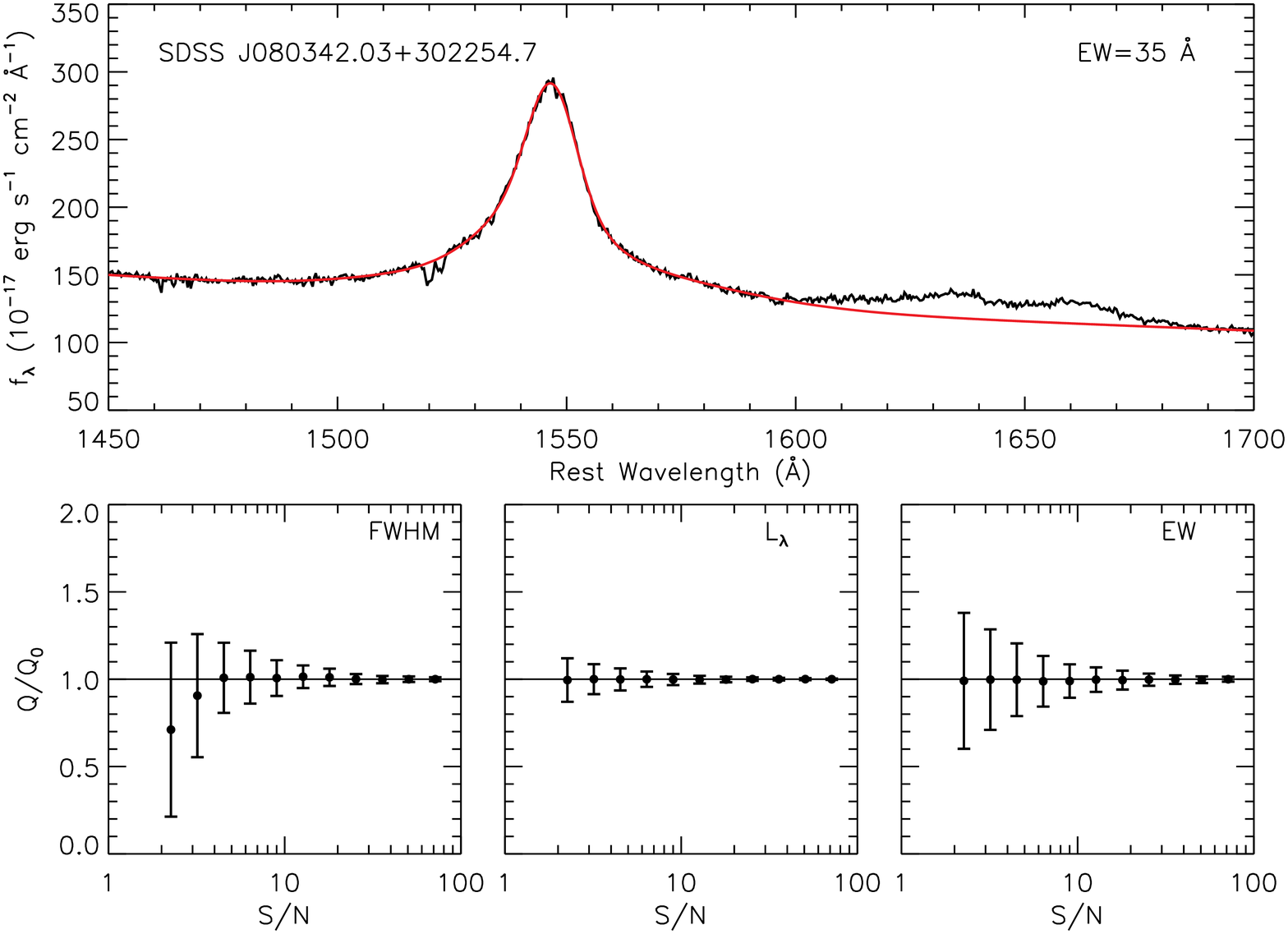}
    \includegraphics[width=0.48\textwidth]{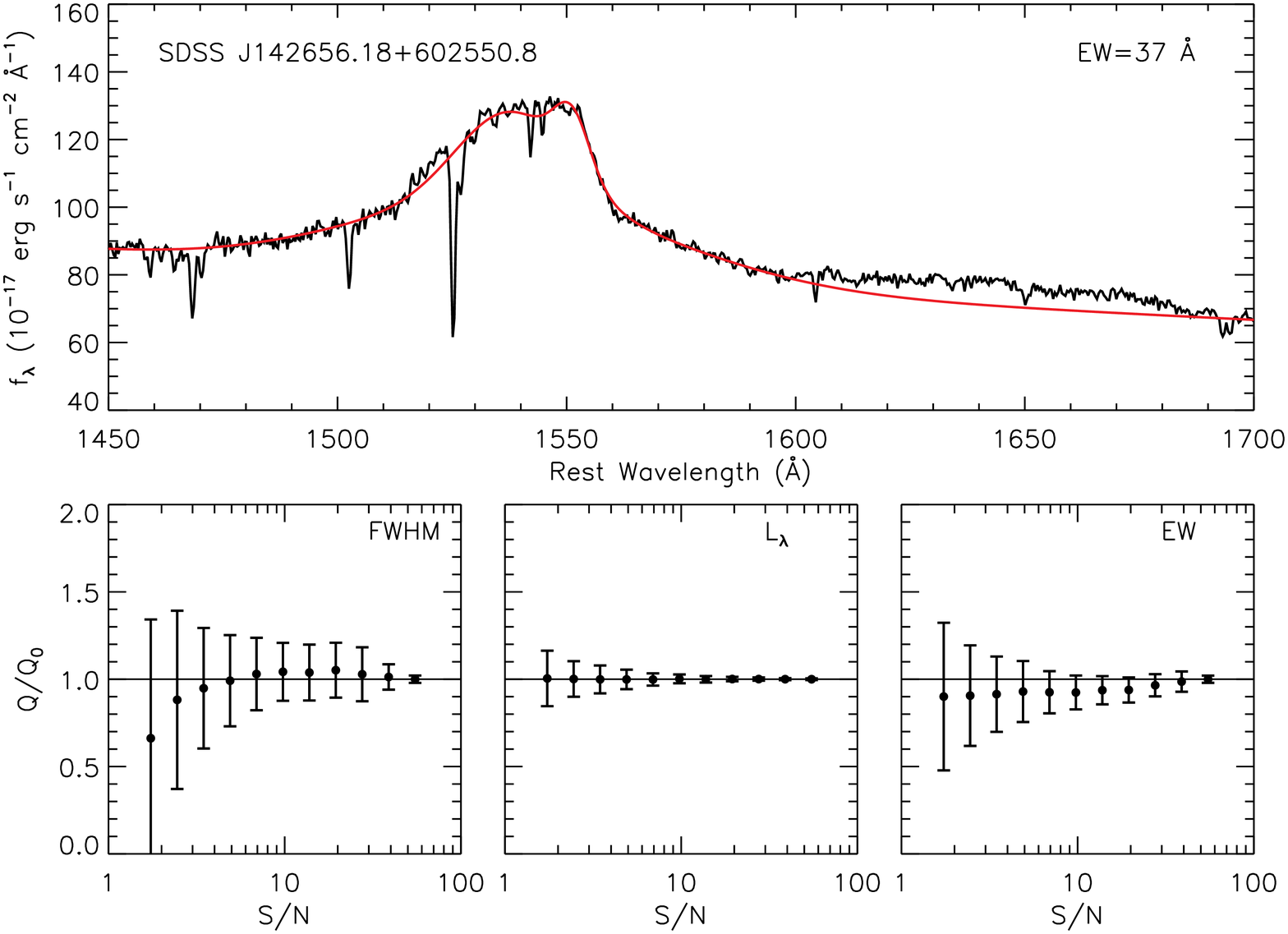}
    \caption{Effects of S/N on the line measurements for \CIV\ for three representative examples. For each object we show
    the actual spectrum (black line) and the best-fit model (red line) in the upper panel. The lower
    three panels show the ratios of the values measured from the degraded spectra to those measured
    from the original spectrum, as functions of S/N; black dots are median values and the error bars indicate the 68\%
    quantile. Note that the He{\tiny II}/O{\tiny III]} complex around 1650\AA\ is not fitted.}
    \label{fig:SN_effect_civ}
\end{figure}

\begin{figure}
  \centering
    \includegraphics[width=0.45\textwidth]{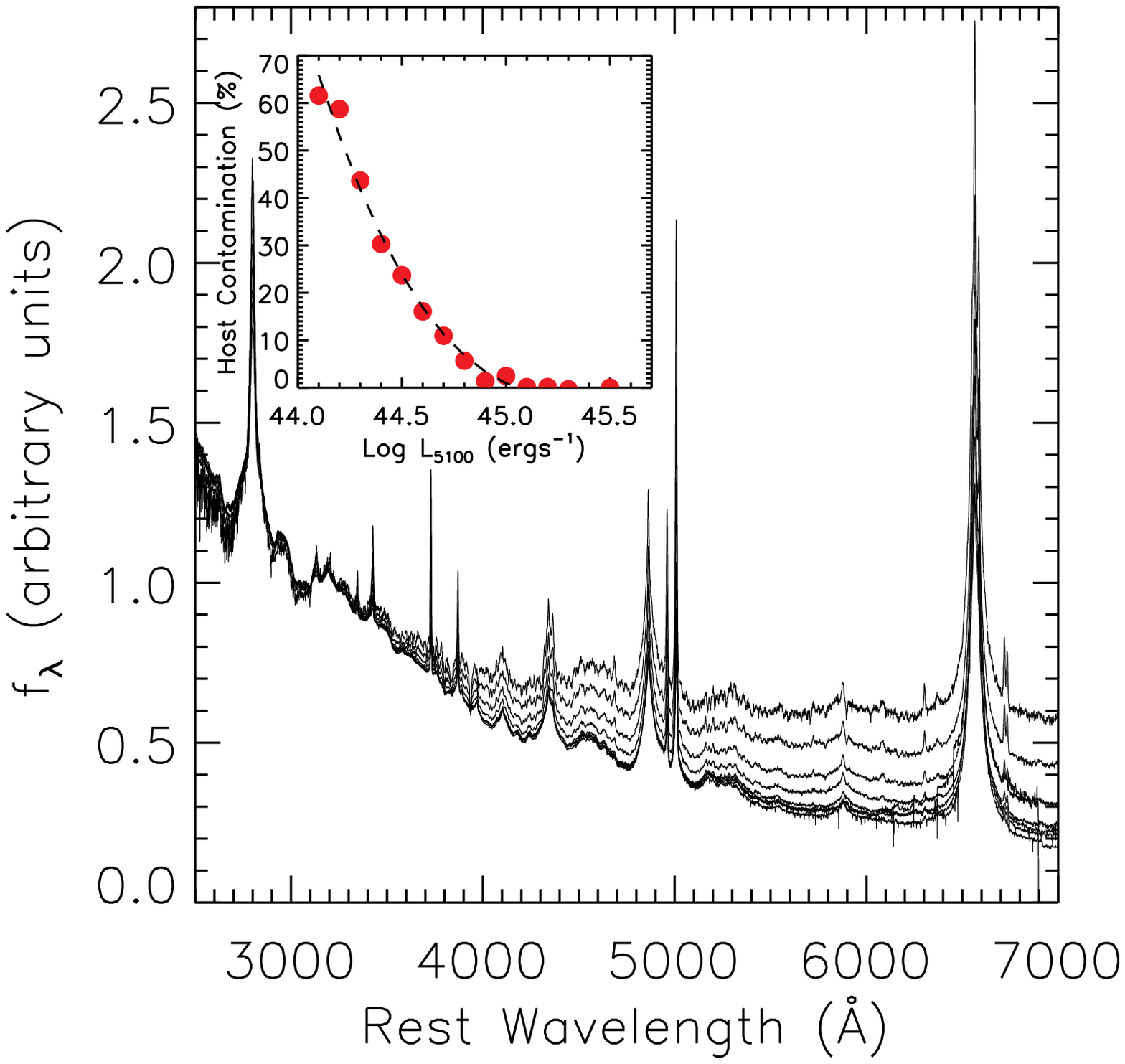}
    \caption{Composite spectra for objects binned in $\log L_{5100}$, normalized at 3000\AA. The flux gradually flattens
    at long wavelength due to increasing host contamination towards fainter quasar luminosities, accompanied
    by the increasing prominence of stellar absorption features and narrow line emission. The inset shows the fractional host contamination at
    $5100$\AA\ assuming that the highest luminosity bin
    ($\log L_{5100}=45.5$) is not affected by host emission and that the intrinsic AGN power-law continuum slope
    does not change over the luminosity range considered. The dashed line in the inset is a polynomial fit (Eqn.\ \ref{eqn:host_fit}).}
    \label{fig:composite_spec}
\end{figure}

Figs.\ \ref{fig:SN_effect_halpha}-\ref{fig:SN_effect_civ} show
several examples of our investigations for \halpha, \hbeta,
\MgII\ and \CIV\ respectively. As expected, decreasing the S/N
increases measurement scatter. In all cases the fitted
continuum is unbiased as S/N decreases. The FWHMs and EWs are
biased by less than $\pm 20\%$ for high-EW objects as S/N is
reduced to as low as $\sim 3$. For low-EW objects, the FWHMs
and EWs are biased low/high by $>20\%$ for S/N$\la 5$. Since
the median EWs for the three lines are $>30$ \AA\ (see
\S\ref{subsec:corr}), we expect that the measurements for most
objects are unbiased to within $\pm 20\%$ down to S/N$\sim 3$.
But for many purposes, it would be more conservative to impose
a cut at S/N$>5$ for reliable measurements.

To estimate the uncertainties in the measured quantities in our
fits, we generate 50 mock spectra by adding Gaussian noise to
the original spectrum using the reported flux density errors,
and fit for those mock spectra with the same fitting routines.
We estimate the measurement uncertainties from the $68\%$ range
(centered on the median) of the distributions of fitting
results of the 50 trials. This approach gives more reasonable
error estimation than using the statistical errors resulting
from the $\chi^2$ fits, in the sense that it not only takes
into account the spectral S/N but also the ambiguity of
subtracting a narrow line component in many cases (for \halpha,
\hbeta\ and \MgII). We estimate the spectral measurement
uncertainties for all the 105,783 quasars in our sample in this
way.

Given the automatic nature and specifics of our fitting recipe,
there will undoubtedly be bad fits for noisy spectra or
peculiar objects. We recommend using the reported measurement
errors to remove suspicious measurements, e.g.,
\texttt{Err}$>0$ \texttt{AND Err}$<$ some threshold. Our
fitting recipe was optimized for the vast majority of quasars
in our sample, and it may fail badly for rare objects with
peculiar continuum and emission line properties. These objects
include severe BALQSOs, disk emitters, and objects such as
J094215.12+090015.8, which has extremely broad lines that
exceed our line fitting range. One should pay attention to our fitting ranges and
these special objects upon usage of the cataloged quantities.
For various reasons, one may want to check the quality
assessment (QA) plot for individual fits to make sure that the
fitting results are robust. Such QA plots are provided along
with this catalog.
%J094215.12+090015.8.

Finally, we note that we imposed a FWHM limit of $1200\,{\rm
km\,s^{-1}}$ for the narrow line components when fitting
\halpha, \hbeta\ and \MgII. This choice is motivated by the
results in \citet{Hao_etal_2005}, but is still somewhat arbitrary;
narrow lines broader than this threshold are not unusual. On
the other hand, if a Gaussian component with FWHM$<1200\,{\rm
km\,s^{-1}}$ can be fit to \halpha, \hbeta\ and \MgII, it will
be considered as a narrow line component and subtracted off.
For these reasons, we urge caution regarding objects for which the
measured narrow line FWHM reaches the $1200\,{\rm km\,s^{-1}}$
limit ($\sim$ a few percent). Upon visual inspection of the QA
plots for these objects, even though the FWHM parameter reaches
the limit, most of the fits still yield reasonable measurements
of total line flux, line centroid and narrow line subtraction.

\subsection{Host galaxy contamination}\label{subsec:host_contam}
For the vast majority of objects in our catalog with $z\ga 0.5$,
host galaxy contamination is negligible. However, for the $z\la 0.5$ low-luminosity quasars in our sample, the continuum luminosity at
restframe $5100$\AA\ may be contaminated by light from the host galaxies.
Unfortunately the spectral quality of the majority of individual objects does not allow a reliable galaxy continuum subtraction. Here we estimate
the effects of host contamination with stacked spectra.

We take all quasars with measurable rest frame $5100$\AA\
continuum luminosity, $L_{5100}$, and bin them on a grid of $\Delta\log
L_{5100}=0.1$ for $\log (L_{5100}/{\rm ergs^{-1}})=44.1-45.5$. Following
\citet{VandenBerk_etal_2001}, we generate geometric mean composite
spectra for objects in each luminosity bin. The composite spectra
are shown in Fig.\ \ref{fig:composite_spec}, where the flux
gradually flattens at long wavelengths due to increasing host
contamination at fainter luminosities. This trend is
accompanied by the increasing prominence of stellar absorption
features and narrow line emission towards fainter luminosities.
The inset shows the fractional host contamination at $5100$\AA\ assuming that the highest luminosity bin ($\log L_{5100}=45.5$) is not affected
by the host and that the intrinsic AGN power-law continuum slope does
not change over the luminosity range considered. The host
contamination is substantial at $\log L_{5100}<44.5$, and becomes
negligible towards higher luminosities. The median value of $\log
L_{5100}$ for quasars in this low-redshift sample is $\sim 44.6$, and therefore the host contamination on average is $\sim 15\%$, which leads to a
$\sim 0.06$ dex overestimation of the $5100$\AA\ continuum
luminosity and thus $\sim 0.03$ dex overestimation of the \hbeta-based
virial masses for the median object. While we do not correct the
measured 5100\AA\ continuum luminosity (and other quantities
depending on it) in the catalog, we provide an empirical fitting
formula of the average host contamination based on the stacked
spectra (dashed line in the inset of Fig.\
\ref{fig:composite_spec}):
\begin{equation}\label{eqn:host_fit}
\frac{L_{\rm 5100,host}}{L_{\rm 5100,QSO}}=0.8052-1.5502x
+ 0.9121x^2-0.1577x^3
\end{equation}
for $x+44\equiv \log (L_{\rm 5100,total}/{\rm erg\,s^{-1}})<45.053$; no correction is
needed for luminosities above this value.

We suspect that host contamination is largely responsible for
the apparent anti-correlation between $L_{5100}$ and spectral slope $\alpha_\lambda$, and the ``negative'' Baldwin effect
\citep[e.g.,][]{Baldwin_1977} between the broad \hbeta\ EW and $L_{5100}$ below $L_{5100}\sim 10^{45}\ {\rm ergs^{-1}}$ (see \S\ref{subsec:corr}).

\subsection{Virial BH masses}\label{subsec:app_vir_mass}

It has become common practice to estimate quasar/AGN BH masses
based on single-epoch spectra (hereafter virial mass in short). This approach assumes that the broad line region (BLR) is virialized, the
continuum luminosity\footnote{We note that in a few extremely radio-loud quasars, the continuum luminosity is significantly boosted by the optical
emission from the jet, which will then lead to overestimation of the BLR size and the virial BH mass \citep[e.g.,][]{Wu_etal_2004}. The fraction of
such objects in our sample is tiny and hence we neglect this detail. But we caution the usage of cataloged virial BH masses for such objects.} is
used
as a proxy for the BLR radius, and the
broad line width (FWHM or line dispersion) is used as a proxy for
the virial velocity.  The virial mass estimate can be expressed
as:
\begin{equation}\label{eqn:virial_estimator}
\log \left({M_{\rm BH,vir} \over M_\odot}\right)
=a+b\log\left({\lambda L_{\lambda} \over 10^{44}\,{\rm
erg\,s^{-1}}}\right)+2\log\left({\rm FWHM\over km\,s^{-1}}\right)\
,
\end{equation}
where the coefficients $a$ and $b$ are empirically calibrated against
local AGNs with RM masses or internally among different lines.  \hbeta, \MgII, \CIV, and their corresponding continuum luminosities are all
frequently
adopted in such virial calibrations. Although it is straightforward to calibrate and use these virial estimators, one must bear in mind the large
uncertainties ($\ga 0.4$ dex) associated with these estimates and the systematics involved in the calibration and usage, which will potentially lead
to significant biases of these BH mass estimates
\citep[e.g.,][]{Collin_etal_2006,Shen_etal_2008b,Marconi_etal_2008,
Denney_etal_2009,Kelly_etal_2009a,Shen_Kelly_2009}.

The virial BH mass calibrations used in this paper are from
\citet[][\hbeta\ and \MgII]{McLure_Dunlop_2004}, \citet[][\hbeta\
and \CIV]{Vestergaard_Peterson_2006}, and
\citet[][\MgII]{Vestergaard_Osmer_2009}. These calibrations have
parameters:
\begin{eqnarray}
(a,b)&=&(0.672,0.61), \qquad {\rm MD04;\ }{\textrm \hbeta\ } \\
(a,b)&=&(0.505,0.62), \qquad {\rm MD04;\ }{\textrm \MgII\ } \\
(a,b)&=&(0.910,0.50), \qquad {\rm VP06;\ }{\textrm \hbeta\ } \\
(a,b)&=&(0.660,0.53), \qquad {\rm VP06;\ }{\textrm \CIV\ } \\
(a,b)&=&(0.860,0.50), \qquad {\rm VO09;\ }{\textrm \MgII\ }
\end{eqnarray}
In using each of these relations we choose the proper FWHM definition
adopted in these calibrations. In order to utilize our new \MgII\
FWHM measurements (e.g., multiple-Gaussian fits with narrow line
subtraction, see \S\ref{subsec:mgii}), we adopt the $b=0.62$ slope in
the BLR radius$-$luminosity relation in \citet{McLure_Dunlop_2004}, and
recalibrate the coefficient $a$ such that the \MgII-based
estimates are consistent with the \hbeta-based (VP06) estimates on
average. We choose this particular slope $b$ because it was re-calibrated in \citet{McLure_Dunlop_2004} using a subsample of reverberation mapping
AGNs that occupy the high-luminosity regime in the local RM AGN sample, arguably better than using the whole RM sample. We choose the VP06 \hbeta\
formula to calibrate our \MgII\ formula because the FWHMs of broad \hbeta\ and \MgII\ were measured in a similar fashion (as opposed to the
single-Gaussian or Lorentzian profile adopted in McLure \& Dunlop 2004). This new \MgII\ calibration is
\begin{equation}
(a,b)=(0.740,0.62), \qquad {\rm S10;\ }{\textrm \MgII\ }\ .
\end{equation}
%We do not utilize other independent calibrations in the literature
%\citep[e.g.,][]{Greene_Ho_2005,McGill_etal_2008,Wang_etal_2009b}, but
%these alternative estimates can be directly computed using our reported
%measurements.

In some cases a particular line may be unavailable in the spectrum, or the continuum is too faint to measure. Below we provide several alternative
empirical recipes to estimate a virial BH mass. These recipes are calibrated using correlations among continuum and emission line properties, and are
{\em only} valid in the average sense. We recommend to use them only when the above estimators are unavailable.

Following \citet{Greene_Ho_2005}, a virial BH mass can be estimated based on the FWHM and luminosity of the broad \halpha\ line. We found a
correlation between the FWHMs (using multiple-Gaussian fits) of the broad \halpha\ and \hbeta\ lines similar to that found in previous work
\citep[e.g.,][]{Greene_Ho_2005,ShenJ_etal_2008}, with the broad \hbeta\ FWHM systematically larger than the broad \halpha\ FWHM:
\begin{eqnarray}
\log\left(\frac{{\rm FWHM_{H\beta}}}{{\rm km\,s^{-1}}}\right)=(&-&0.11\pm0.03)\nonumber\\
&+&(1.05\pm0.01)\log\left(\frac{{\rm FWHM_{H\alpha}}}{{\rm
km\,s^{-1}}}\right)\,
\end{eqnarray}
where the slope and intercept are determined using the BCES bisector linear regression estimator (e.g., Akritas \& Bershady 1996) for a sample of
$\sim 2400$ quasars with both \halpha\ and \hbeta\ FWHM measurements and FWHM errors less than $500\,{\rm kms^{-1}}$.
Our continuum luminosities at 5100\,\AA\ have a narrow dynamical range and suffer from host contamination at the low-luminosity end, hence instead of
fitting a new relation using our measurements, we adopt the relation between 5100\,\AA\ continuum luminosity and \halpha\ line luminosity in
\citet[][their eqn.\ 1]{Greene_Ho_2005}. The virial mass estimator based on \halpha\ therefore reads
\begin{eqnarray}\label{eqn:vir_halpha}
\log \left({M_{\rm BH,vir} \over M_\odot}\right)_{\rm H\alpha}=&&0.379 + 0.43\log\left(\frac{L_{\rm H\alpha}}{\rm 10^{42}\, erg\,s^{-1}}\right)\nonumber\\
 &+& 2.1\log\left(\frac{{\rm FWHM_{H\alpha}}}{{\rm km\,s^{-1}}}\right)\ ,
\end{eqnarray}
where $L_{\rm H\alpha}$ is the total \halpha\ line luminosity.
For quasars in our sample, Eqn.\ (\ref{eqn:vir_halpha}) yields
virial BH masses consistent with the VP06 \hbeta\ results, with
a mean offset $\sim 0.08$ dex and a dispersion $\sim 0.18$ dex.

Similarly, we can substitute the continuum luminosity in the
above recipes for \hbeta, \MgII\ and \CIV\ with the luminosity
of the particular line used, given that for broad line quasars
the line luminosity correlates with the continuum luminosity to
some extent. We determine these correlations using subsamples
of quasars for which both luminosities were measured with an
uncertainty $<0.03$ dex. For the same reason as \halpha, we do
not fit a new relation between the \hbeta\ line luminosity and
continuum luminosity at 5100\,\AA, and we refer to eqn.\ (2) of
\citet{Greene_Ho_2005} for such a relation. The following
relations are determined again using the BCES linear regression
estimator (for both the bisector fit and the $Y|X$ fit, where
the latter refers to ``predict $Y$ as a function of $X$''),
where the line luminosity refers to the total line
luminosity\footnote{For \MgII, the difference between the total
line luminosity and the broad line luminosity is small enough
such that it essentially makes no difference in the linear
regression results when we use the broad line luminosity
instead.}:
\begin{eqnarray}
\log \left(\frac{L_{\rm MgII}}{\rm erg\,s^{-1}}\right)&=&(2.22\pm0.09)
\\ &+& (0.909\pm0.002)\log \left(\frac{L_{3000}}{\rm erg\,s^{-1}}\right)\ ,\quad
{\rm bisector}\ ;\nonumber\\
\log \left(\frac{L_{3000}}{\rm erg\,s^{-1}}\right)&=&(1.22\pm0.11)\\
&+&(1.016\pm0.003)\log \left(\frac{L_{\rm MgII}}{\rm erg\,s^{-1}}\right)\ ,\quad {\rm
(Y|X)}\ ,\nonumber
\end{eqnarray}
where the scatter of this correlation is $\sim 0.15$ (0.16) dex for $\sim 44,000$ quasars, and
\begin{eqnarray}
\log \left(\frac{L_{\rm CIV}}{\rm erg\,s^{-1}}\right)&=&(4.42\pm0.27)\\
&+&(0.872\pm0.006)\log \left(\frac{L_{1350}}{\rm erg\,s^{-1}}\right)\ ,\quad {\rm
bisector}\ ;\nonumber\\
\log \left(\frac{L_{1350}}{\rm erg\,s^{-1}}\right)&=&(7.66\pm0.41)\\
&+&(0.863\pm0.009)\log \left(\frac{L_{\rm CIV}}{\rm erg\,s^{-1}}\right)\ ,\quad {\rm
(Y|X)}\ ,\nonumber
\end{eqnarray}
where the scatter of this correlation is $\sim 0.18$ (0.2) dex for $\sim 10,000$ quasars. Using our measurements, one can also estimate these
correlations with other linear regression algorithms. If we substitute $\log L_{3000}$ and $\log L_{1350}$ using Eqns.\ (12) and (14) in the \MgII\
and \CIV\ estimators, we obtain virial BH masses consistent with the original recipes with no mean offset ($<0.02$ dex) and negligible scatter ($\sim
0.1$ dex).

%We re-emphasize that in these calibrations different line-fitting
%procedures can lead to difference in FWHM as large as $\sim 0.2$
%dex for individual objects, hence one should choose the proper
%FWHM measure when using a particular calibration.

\begin{figure*}
  \centering
    \includegraphics[width=0.45\textwidth]{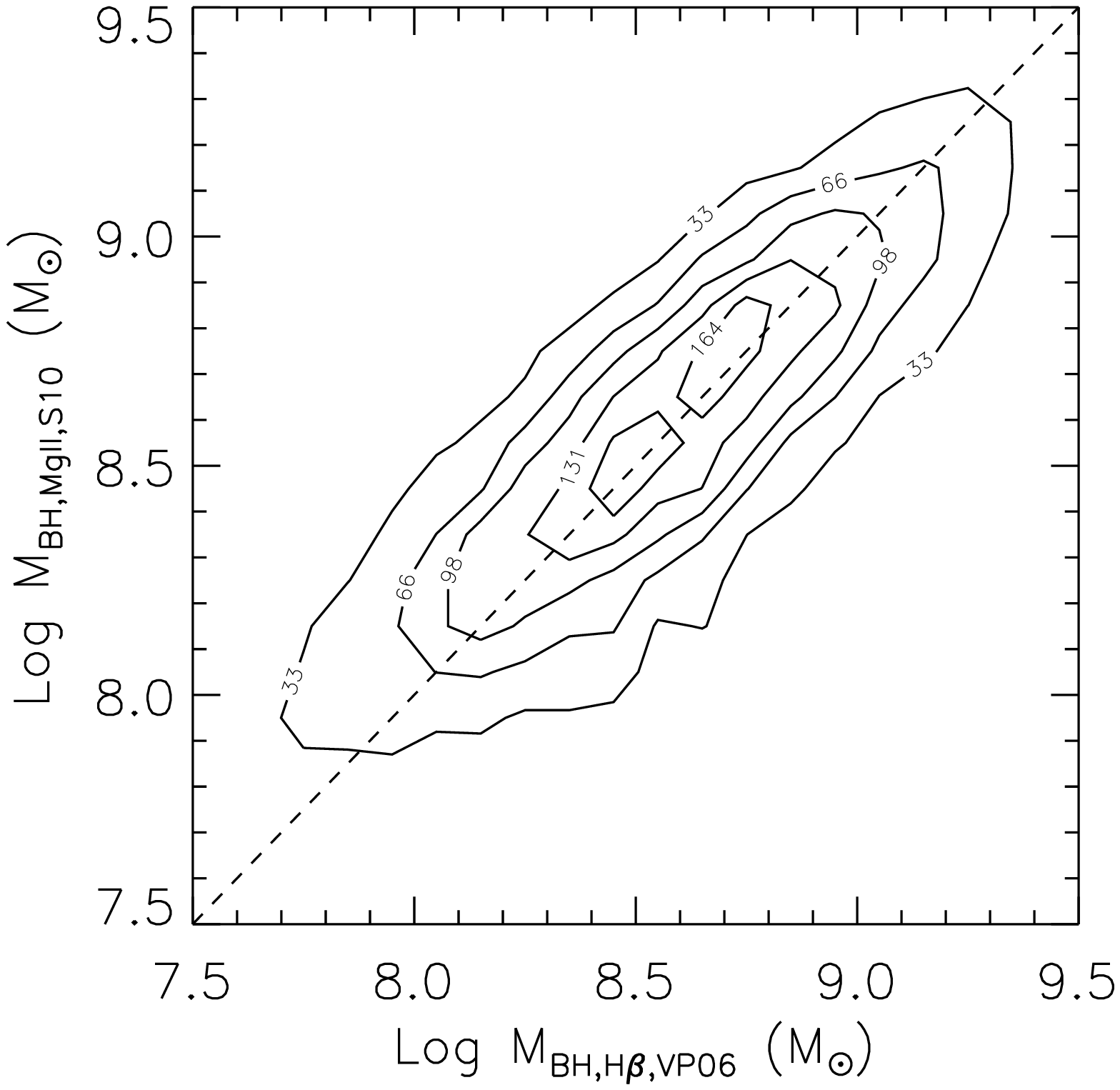}
    \includegraphics[width=0.45\textwidth]{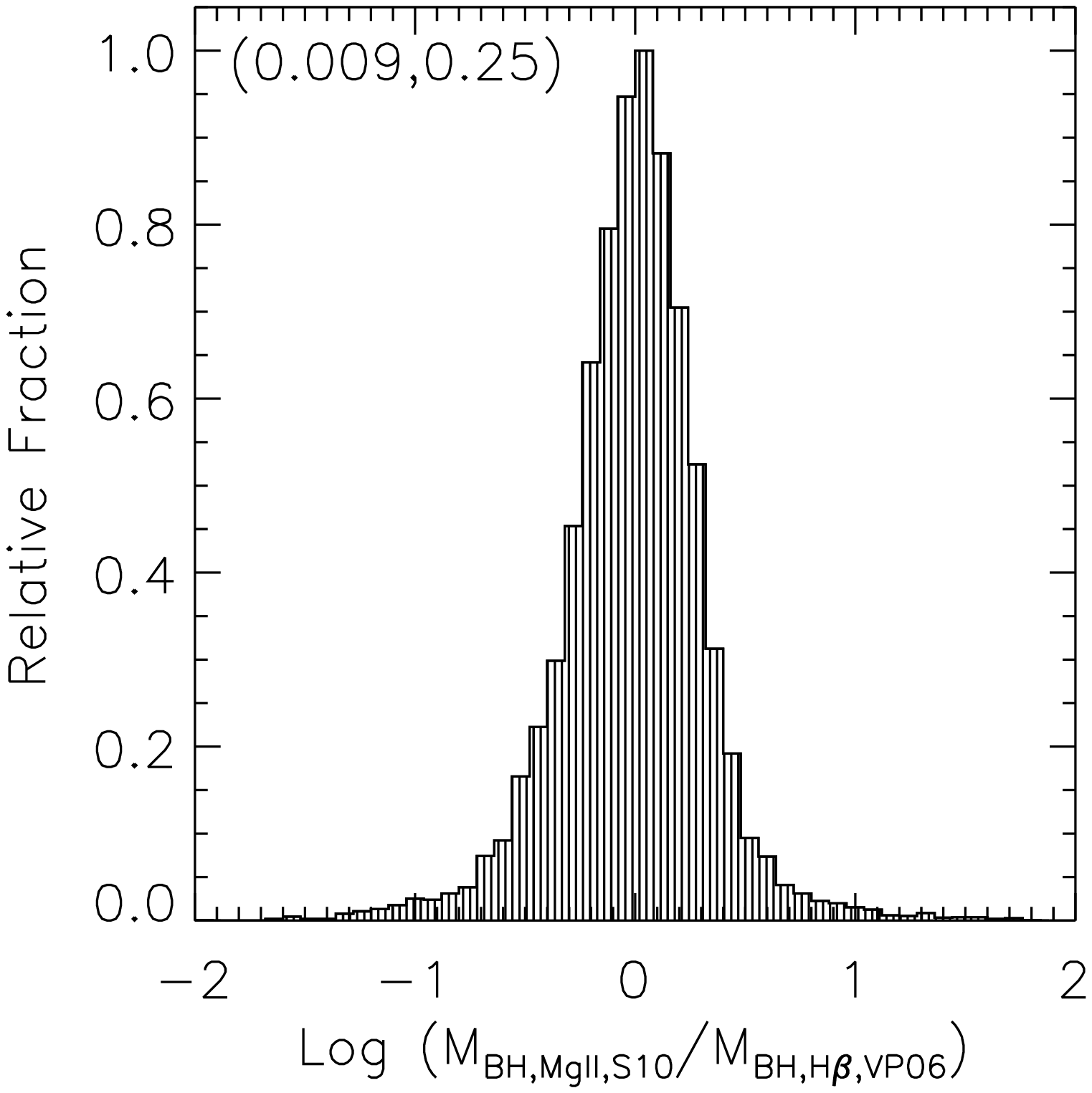}
    \includegraphics[width=0.45\textwidth]{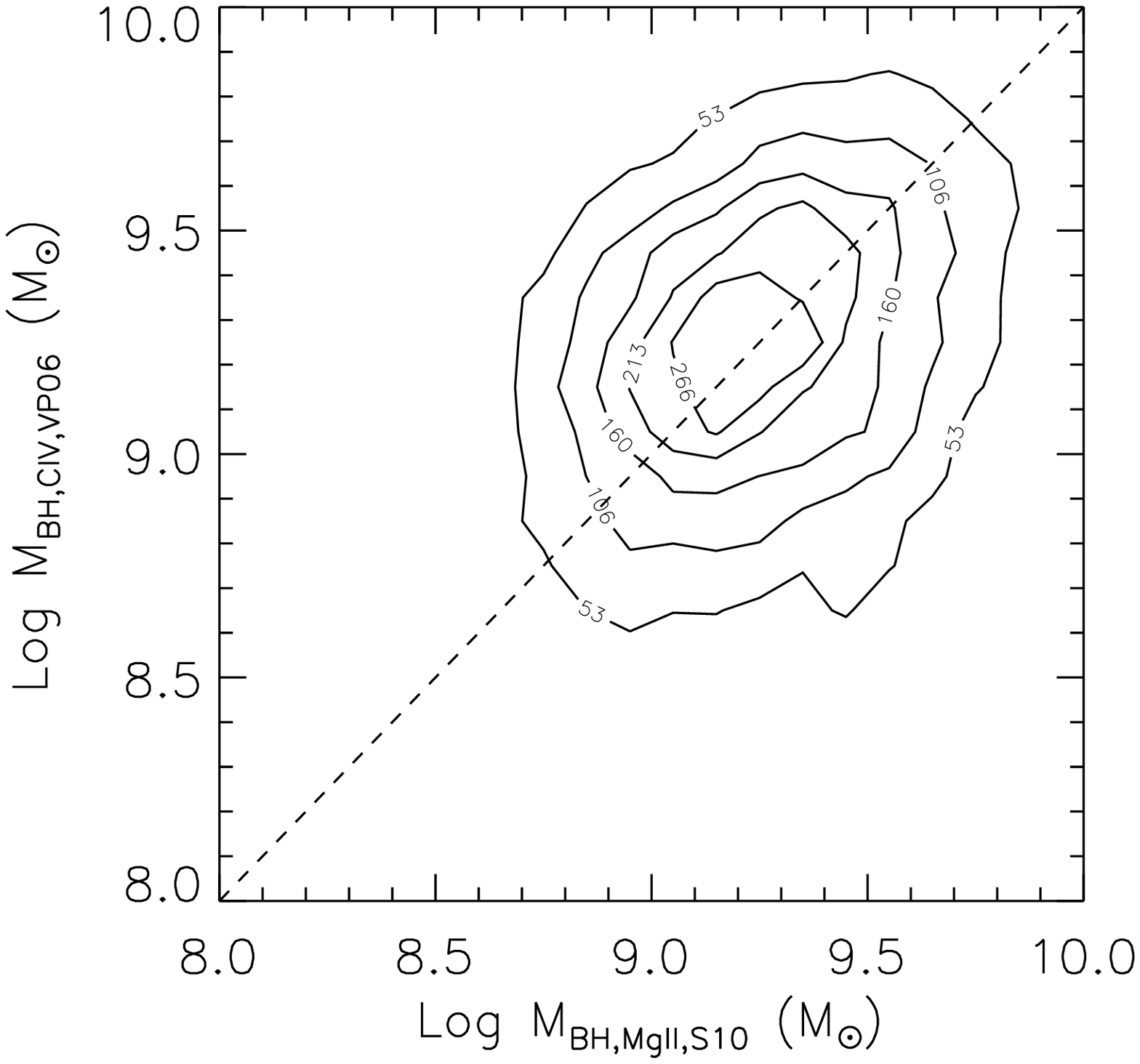}
    \includegraphics[width=0.45\textwidth]{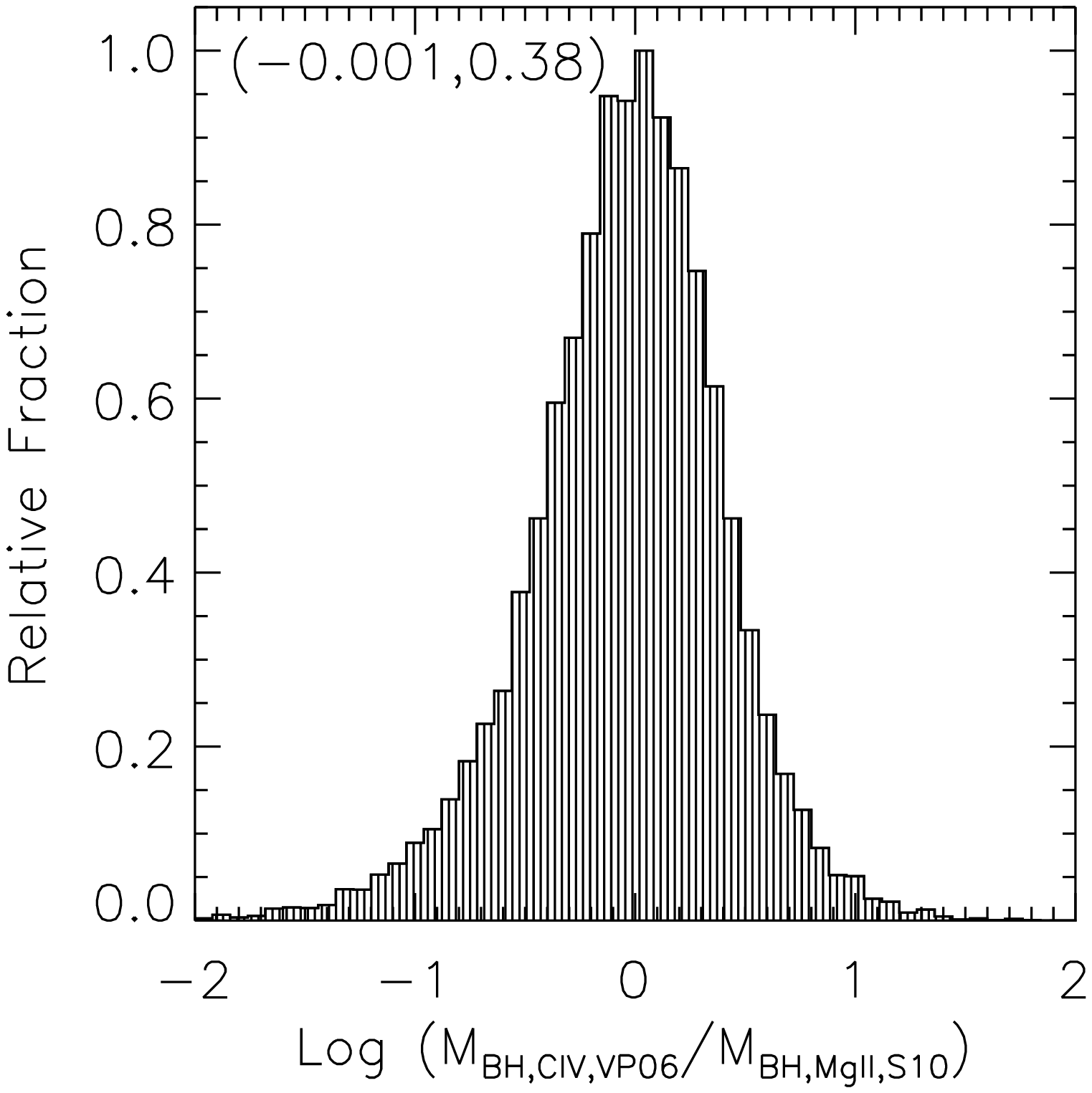}
    \caption{Comparison of virial masses between two different line estimators for the subset of quasars in our sample for which both line estimates
    are available and the median line S/N per pixel $>6$. The left panels are one-to-one plots, where the contours
    are local point density contours estimated with a grid size of $\Delta=0.1$ on both axes. The right panels show the distribution of mass ratios
    between
    two lines, and the mean and $1\sigma$ from a Gaussian fit to the distribution are indicated in
    the top-left corner. Our new \MgII\ recipe gives consistent virial mass estimates as the VP06 \hbeta\ recipe (by design), and it also yields
    consistent virial mass estimates as \CIV\ (VP06). These comparison results are similar to those in \citet[][fig.\ 6]{Shen_etal_2008b}, where we
    compared between \hbeta\ (MD04), \MgII\ (MD04) and \CIV\ (VP06).}
    \label{fig:com_vir_mass}
\end{figure*}

%Fig.\ \ref{fig:com_vir_mass} compares the virial mass estimates
%between different lines and different calibrations for objects
%with median line S/N$>6$. We have used the corresponding FWHM
%measurements for the specific virial calibration as described in
%\S\ref{subsec:cat}.
There are systematic differences among different versions of
virial calibrations. For instance, the calibrations for \hbeta\
and \MgII\ in \citet{McLure_Dunlop_2004} used the old RM masses
and virial coefficient, while those in \citet{Vestergaard_Peterson_2006} and \citet{Vestergaard_Osmer_2009} used the updated RM masses and
virial coefficient \citep{Onken_etal_2004}. Significant uncertainties of the virial coefficient still remain
\citep[e.g.,][]{Woo_etal_2010,Graham_etal_2011}. Moreover, different versions of virial calibration for the same line have different dependence on
luminosity, and they usually measure the line FWHM differently
(even though occasionally different approaches to measure the FWHM
yield the same value during the multi-parameter fits), or prefer an alternative proxy (e.g., line dispersion) for the virial velocity
\citep[e.g.,][]{Collin_etal_2006,Rafiee_Hall_2010}. Currently there is no consensus on which version of calibration is better. It is
important to explore various systematics with RM AGN samples and
statistical quasar samples to determine which is the best approach
to estimate quasar BH masses with the virial technique, and this
is work in progress
\citep[e.g.,][]{Onken_Kollmeier_2008,Denney_etal_2009,Wang_etal_2009b,Rafiee_Hall_2010}. In the mean time, there is strong need to increase the
sample
size and representativity of AGNs with RM measurements, which anchor these single-epoch virial estimators.

Here we simply settle on a fiducial virial mass estimate: we
use \hbeta\ (VP06) estimates for $z<0.7$, \MgII\ (S10)
estimates for $0.7\le z<1.9$ and \CIV\ (VP06) estimates for
$z\ge 1.9$. Fig.\ \ref{fig:com_vir_mass} shows the comparison
between these virial estimates between two lines for the subset
of quasars for which both line estimates are available and the
median line S/N$>6$. There is negligible mean offset ($<0.01$
dex) between these virial estimates (half by design), which
motivated our choice of these three calibrations. However, as
noted in \citet{Shen_etal_2008b}, there is a strong trend of
decreasing the ratio of $\log (M_{\rm BH}^{\rm Mg{\tiny
II}}/M_{\rm BH}^{\rm C\tiny{IV}})$ with increasing \CIV-\MgII\
blueshifts, indicating a possible non-virialized component in
\CIV. \citet{Richards_etal_2011} further developed a unified
picture in which the \CIV\ line has both a non-virial wind
component and a traditional virial component, and it is
plausible that the BH mass scaling relation based on \CIV\ is
only relevant for objects dominated by the virial component.

\subsection{The spectral catalog}\label{subsec:cat}

We have tabulated all the measured quantities from the spectral
fitting in the online catalog of this paper,
along with other properties\footnote{Note that in
\citet{Shen_etal_2008b} we only reported high-quality measurements
with median S/N$>6$ and a reduced $\chi^2<5$ for a single-Gaussian
fit to the line; here we retain all measurements for
completeness.}. The current compilation extends our earlier SDSS Data Release 5 (DR5) compilation \citep{Shen_etal_2008b} by including the post-DR5
quasars, as well as measurements based on new multiple-Gaussian fits to the lines (as discussed above). The format of the catalog is described in
Table 1.
Objects are in the same order as the DR7 quasar catalog in
\citet{Schneider_etal_2010}. Below we describe the specifics of
the cataloged quantities. The SDSS terminology can be found on the SDSS
website\footnote{http://www.sdss.org/dr7/}. Flux measurements were corrected neither for intrinsic extinction/reddening, nor for host contamination.
We only report FWHM and velocity shift values for detectable line components (i.e., the fitted line flux is non-zero), except for cases where the
line
FWHM and velocity offset can be inferred from other lines (such as the narrow \hbeta\ line, whose FWHM and velocity offset are tied to those of the
\OIII\ doublet).

\begin{enumerate}

\item SDSS DR7 designation: {\em hhmmss.ss+ddmmss.s} (J2000.0;
truncated coordinates)

\item[2-4.] RA and DEC (in decimal degrees, J2000.0), redshift. Here the redshifts are taken from the DR7 quasar catalog
    \citep{Schneider_etal_2010}. \citet{Hewett_Wild_2010} provided improved redshifts for SDSS quasars. These improved redshifts are particularly
    useful for generating coadded spectra, but the cataloged DR7 redshifts are fine for most of the purposes considered here.

\item[5-7.] Spectroscopic plate, fiber and Modified Julian date (MJD): the combination of
plate-fiber-MJD locates a particular spectroscopic observation in
SDSS. The same object can be observed more than once with
different plate-fiber-MJD combinations
either on a repeated plate (same plate and fiber numbers but
different MJD number), or on different plates. The DR7 quasar catalog typically lists the spectroscopic observation with the highest S/N.

\item[8.] TARGET\_FLAG\_TARGET: the target selection flag (TARGET version).

\item[9.] $N_{\rm spec}$: number of spectroscopic observations. While we only used the default
spectrum in our spectral fitting, this flag indicates if there are multiple spectroscopic observations
for each object.

\item[10.] Uniform flag. 0=not in the uniform sample\footnote{For more details regarding the uniform sample selection and its sky coverage, see,
    e.g., \citet{Richards_etal_2002a,Richards_etal_2006a,Shen_etal_2007b}.}; 1=uniformly selected using the target selection algorithm in
    \citet{Richards_etal_2002a}, and flux
limited to $i=19.1$ at $z<2.9$ or $i=20.2$ at $z>2.9$; 2=selected
by the \texttt{QSO\_HiZ} branch only in the uniform target
selection \citep{Richards_etal_2002a} and with measured
spectroscopic redshift $z<2.9$ and $i>19.1$. Objects with uniform
flag=2 are selected by the uniform quasar target algorithm, but
should not be included in statistical studies; the fraction of
such uniform objects is low ($<1\%$; red dots in Fig.\ \ref{fig:mi_dist}).

\item[11.] $M_i(z=2)$: absolute $i$-band magnitude in the current
cosmology, $K$-corrected to $z=2$ following\footnote{The $K$-corrections here include both
continuum $K$-correction and emission-line $K$-correction \citep{Richards_etal_2006a}; while the cataloged absolute magnitudes in
\citet{Schneider_etal_2010} were $K$-corrected for continuum only, assuming a power-law continuum.}
\citet{Richards_etal_2006a}.

\item[12-13.] Bolometric luminosity $L_{\rm bol}$ and its error: computed from $L_{5100}$ ($z<0.7$), $L_{3000}$ ($0.7\le z<1.9$), $L_{1350}$
    ($z\ge 1.9$) using the spectral fits and bolometric
corrections\footnote{The SEDs for individual quasars show significant scatter \citep[e.g.,][]{Richards_etal_2006b},
so the adopted bolometric corrections are only appropriate in the average sense. Some authors suggest to remove
the IR bump in the SED in estimating the bolometric corrections \citep[e.g.,][]{Marconi_etal_2004}, where the
IR radiation is assumed to come from the reprocessed UV radiation. This will generally reduce the bolometric corrections by
about one third.
%Integrating the mean SED in \citet{Richards_etal_2006b} from 1 \micron\ to 10 kev for the bolometric luminosity we obtain:
% ${\rm BC}_{5100}=5.47$, ${\rm BC}_{3000}=3.04$ and ${\rm BC}_{1350}=2.25$.
However, we are not correcting for intrinsic extinction of the flux, and we did not subtract emission line flux in the Richards et~al. composite
SED when estimating the bolometric corrections, hence using our fiducial bolometric corrections will not overestimate the bolometric luminosity
significantly.} ${\rm BC}_{5100}=9.26$, ${\rm BC}_{3000}=5.15$ and ${\rm BC}_{1350}=3.81$ from the composite SED in \citet{Richards_etal_2006b}.

\item[14.] BAL flag: 0=nonBALQSO or no wavelength coverage; 1=\CIV\ HiBALQSO; 2=\MgII\ LoBALQSO; 3=both 1 and 2. The LoBALQSO selection is
very incomplete as discussed in \S\ref{sec:sample}.

\item[15.] FIRST radio flag: $-1$=not in FIRST footprint; 0=FIRST
undetected; 1=core dominant; 2=lobe dominant \citep[for details, see][and discussion in \S\ref{sec:sample}]{Jiang_etal_2007a}. Note these two
classes of radio morphology do not necessarily correspond to the FR I and FR II types (see Lin et al.\ 2010 for more details).

\item[16-17.] {\em Observed} radio flux density at rest-frame $6$ cm
$f_{\rm 6cm}$ and optical flux density at rest-frame $2500$ \AA\
$f_{2500}$. Both are in the observed frame.

\item[18.] Radio loudness $R\equiv f_{\rm 6cm}/f_{2500}$.

\item[19-24.] $L_{5100}$, $L_{3000}$, $L_{1350}$ and their errors: continuum
luminosity at $5100$ \AA, $3000$ \AA\ and $1350$ \AA, measured
from the spectral fits. No correction for host contamination is
made (see discussion in \S\ref{subsec:host_contam}).

\item[25-30.] Line luminosity, FWHM, restframe equivalent width and their errors for the broad \halpha\ component.

\item[31-36.] Line luminosity, FWHM, restframe equivalent width and their errors for the narrow \halpha\ component.

%\item[27-28.] line luminosity and equivalent width for narrow
%\NIIa.

\item[37-40.] Line luminosity, restframe equivalent width and their errors for narrow \NIIb.

\item[41-44.] Line luminosity, restframe equivalent width and their errors for narrow \SIIa.

\item[45-48.] Line luminosity, restframe equivalent width and their errors for narrow \SIIb.

\item[49-50.] 6000-6500 \AA\ iron restframe equivalent width and its error.

\item[51-52.] Power-law slope $\alpha_\lambda$ and its error for the continuum fit for \halpha.

\item[53-54.] Number of good pixels and median ${\rm S/N}$ per
pixel for the \halpha\ region (6400-6765 \AA).

\item[55.] Reduced $\chi^2$ for the \halpha\ line fit; $-1$ if not
fitted.

\item[56-61.] Line luminosity, FWHM, restframe equivalent width and their errors for the broad \hbeta\ component.

\item[62-67.] Line luminosity, FWHM, restframe equivalent width and their errors for the narrow \hbeta\ component.

\item[68.] FWHM of broad \hbeta\ using a single Gaussian fit
\citep{Shen_etal_2008b}.

\item[69-72.] Line luminosity, restframe equivalent width and their errors for \OIIIa.

\item[73-76.] Line luminosity, restframe equivalent width and their errors for \OIIIb.

\item[77-78.] 4435-4685 \AA\ iron restframe equivalent width and its error.

\item[79-80.] Power-law slope $\alpha_\lambda$ and its error for the continuum fit for \hbeta.

\item[81-82.] Number of good pixels and median ${\rm S/N}$ per
pixel for the \hbeta\ region (4750-4950 \AA).

\item[83.] Reduced $\chi^2$ for the \hbeta\ line fit; $-1$ if not
fitted.

\item[84-89.] Line luminosity, FWHM, restframe equivalent width and their errors for the whole \MgII\ profile.

\item[90-95.] Line luminosity, FWHM, restframe equivalent width and their errors for the broad \MgII\ profile.

\item[96.] FWHM of broad \MgII\ using a single Gaussian fit
\citep{Shen_etal_2008b}.

\item[97-98.] 2200-3090 \AA\ iron restframe equivalent width and its error.

\item[99-100.] Power-law slope $\alpha_\lambda$ and its error for the continuum fit for \MgII.

\item[101-102.] Number of good pixels and median ${\rm S/N}$ per
pixel for the \MgII\ region (2700-2900 \AA).

\item[103.] Reduced $\chi^2$ for the \MgII\ line fit; $-1$ if not
fitted.

\item[104-109.] Line luminosity, FWHM, restframe equivalent width and their errors for the whole \CIV\ profile.

%\item[69.] 1400-1700 \AA\ iron equivalent width.

\item[110-111.] Power-law slope $\alpha_\lambda$ and its error for the continuum fit for \CIV.

\item[112-113.] Number of good pixels and median ${\rm S/N}$ per
pixel for the \CIV\ region (1500-1600 \AA).

\item[114.] Reduced $\chi^2$ for the \CIV\ line fit; $-1$ if not
fitted.

\item[115-126.] Velocity shifts (and their errors) relative to the systemic redshift \citep[cataloged in][]{Schneider_etal_2010}
for broad \halpha, narrow \halpha, broad \hbeta, narrow \hbeta,
broad \MgII, and \CIV. The velocity shifts for the broad
lines are measured from the peak of the multiple-Gaussian model fit to the broad component\footnote{The velocity shifts of the broad lines
measured from the centroid of a single Gaussian fit to the line on average are consistent with those using the peak of the multiple-Gaussian fit
with negligible mean offset, but they can differ (typically by $\la 200\ {\rm kms^{-1}}$) for individual objects.}. Recall that the velocity
shifts of narrow lines were tied together during spectral fits. These velocity shifts can be used to compute the relative velocity offsets
between
two lines for the same object, such as the \CIV-\MgII\ blueshift, but should {\em not} be interpreted as the velocity shifts from the restframe
of
the host galaxy due to uncertainties in the systemic redshift. Positive values indicate blueshift and negative values indicate redshift; value of
$3\times 10^5$ indicates an unmeasurable quantity.

\item[127-138.] Virial BH masses using calibrations of \hbeta\
(MD04), \hbeta\ (VP06), \MgII\ (MD04), \MgII\ (VO09), \MgII\ (S10)
and \CIV\ (VP06). The definitions of the acronym names of each calibration can be found in \S\ref{subsec:app_vir_mass}. Zero value indicates an
unmeasurable quantity. We use FWHMs
from a single Gaussian fit to the broad component for \hbeta\
(MD04) and \MgII\ (MD04); FWHMs from the multiple-Gaussian fit to
the broad \hbeta\ for \hbeta\ (VP06); FWHMs from the
multiple-Gaussian fit to the entire \MgII\ and \CIV\ lines for
\MgII\ (VO09) and \CIV\ (VP06) respectively; FWHMs from the
multiple-Gaussian fit to the broad \MgII\ line for \MgII\ (S10).
See \S\ref{sec:spec_measure} for details.

\item[139.] The adopted fiducial virial BH mass if more than one
estimate is available. See detailed discussion in
\S\ref{subsec:app_vir_mass}.

\item[140.] The measurement uncertainty of the adopted fiducial virial BH mass, propagated from the measurement uncertainties of continuum
    luminosity and FWHM. Note that this uncertainty includes neither the statistical uncertainty ($\ga 0.3-0.4$ dex) from virial mass
    calibrations, nor the systematic uncertainties with these virial BH masses.

\item[141.] Eddington ratio computed using the fiducial virial BH
mass.

%\item[??] Spectral indices?

\item[142.] Special interest flag. This is a binary flag: bit\#0 set=disk emitters with high confidence (the vast majority are selected based on
    the Balmer lines); bit\#1 set=disk emitter candidates; bit\#2 set=double-peaked \OIIIab\ lines. These flags were set upon visual inspection
    of
    all $z<0.89$ quasars in the catalog. In particular, disk emitter candidates (bit\#1=1) are those with asymmetric broad Balmer line profile or
    systematic velocity shifts from the narrow lines; while those with high confidence (bit\#0=1) show unambiguous double-peaked (or highly
    asymmetric) broad line profile or large velocity offsets between the broad and narrow lines. Fig.\ \ref{fig:disk_emitter} shows two examples
    of disk emitters with high confidence. We call these objects ``disk emitters'' even though some of them may be explained by alternative
    scenarios, such as a close SMBH binary \citep[e.g., see discussion in][]{Shen_Loeb_2009}.

\end{enumerate}

\begin{figure}
  \centering
    \includegraphics[width=0.45\textwidth]{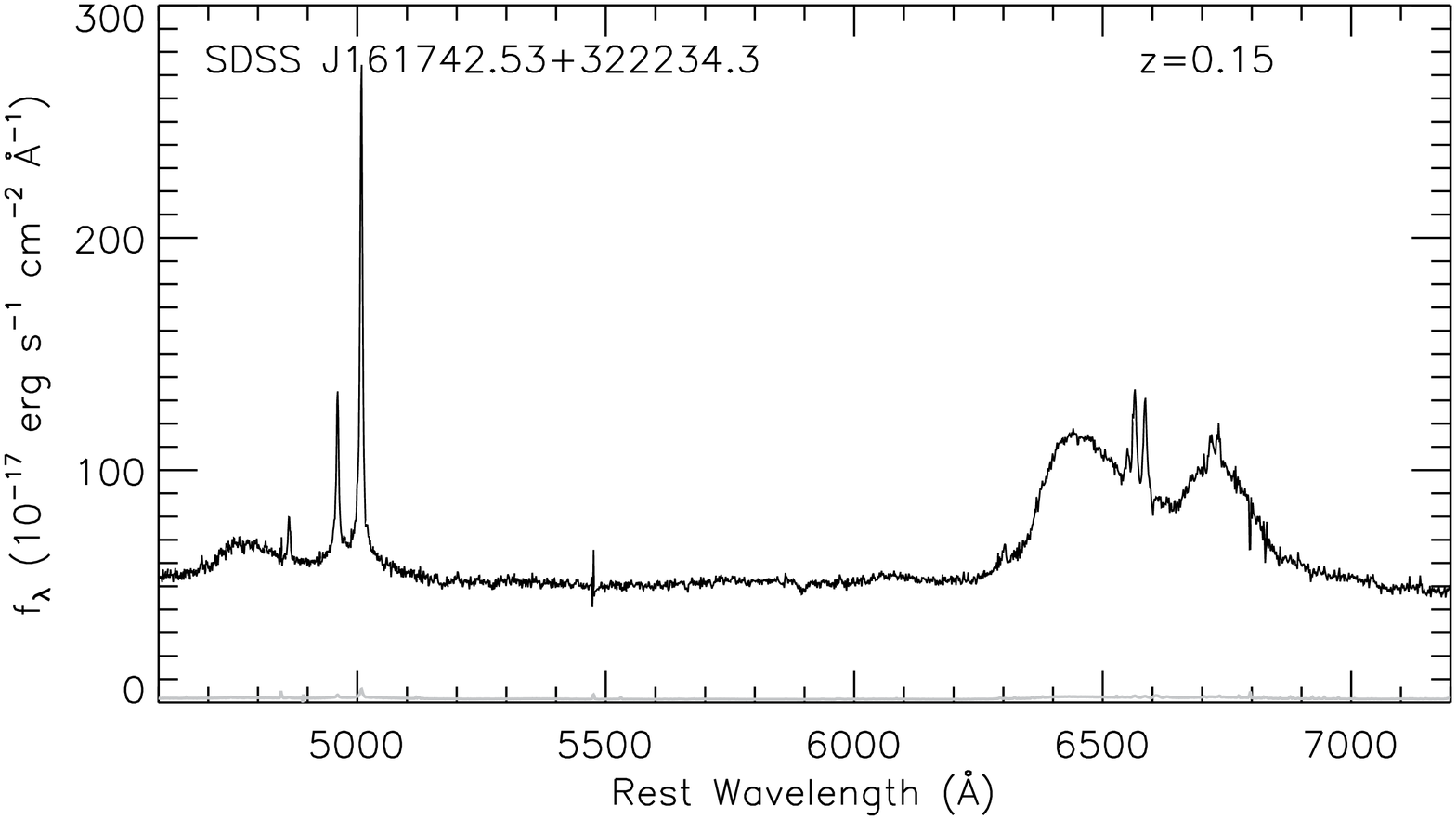}
    \includegraphics[width=0.45\textwidth]{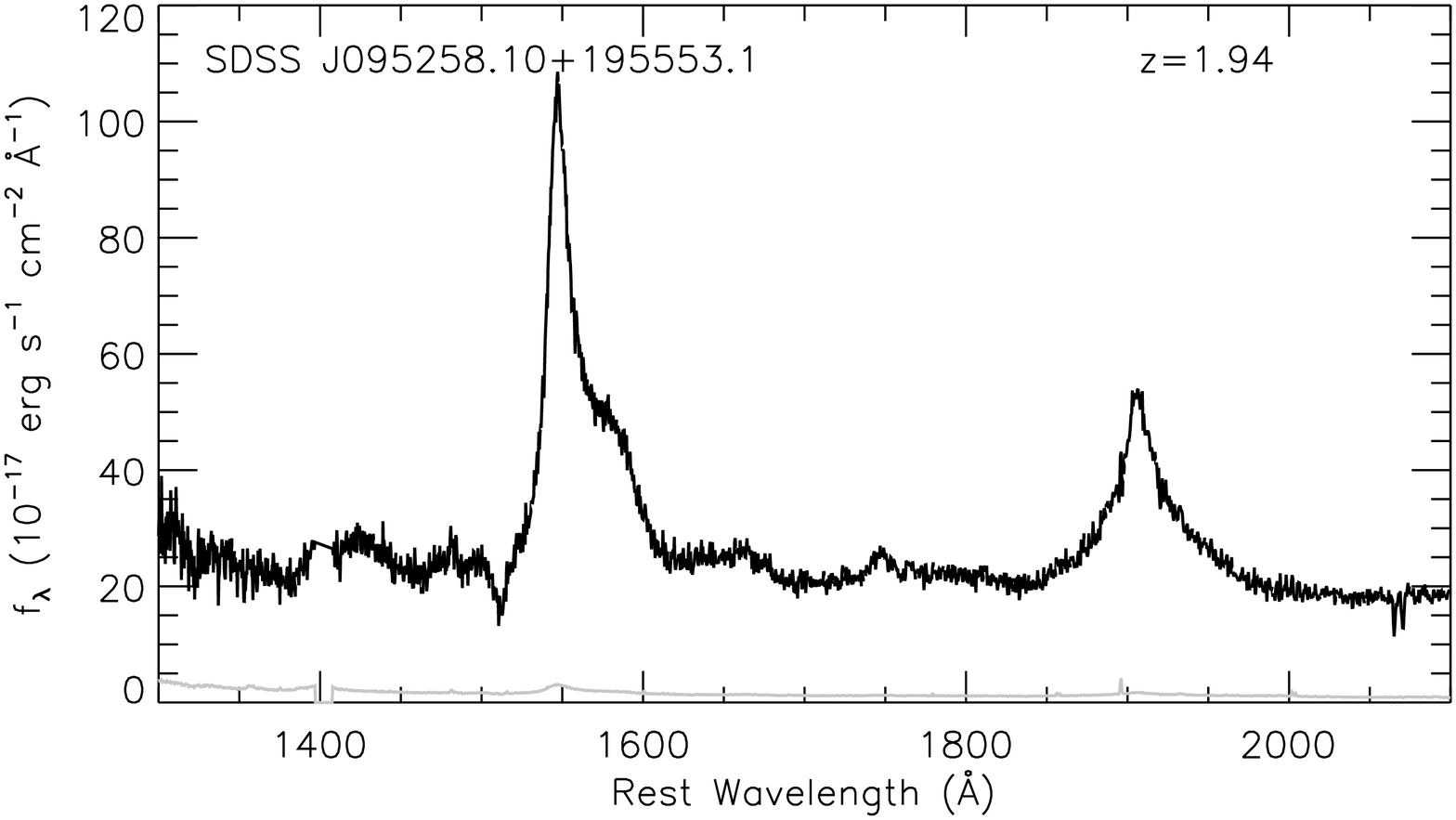}
    \caption{Two examples of disk emitters with high confidence (bit\#0=1). {\em Upper:} a Balmer disk emitter at $z=0.15$. {\em Bottom:} a possible
    CIV disk emitter at $z=1.94$.}
    \label{fig:disk_emitter}
\end{figure}

\section{Applications}\label{sec:app}

The spectral measurements described above can be used to study the
statistical properties of broad line quasars. Here we briefly discuss
some applications of this spectral catalog.

%{\bf Describe some caveats in the line fitting. 1) Since the fits
%are multi-parameters, sometimes the fits are incorrect for the
%narrow lines (i.e., very large line luminosity) but OK for the
%broad lines, or over-corrected iron template and hence incorrect
%continuum fits (I will exclude them using
%``add\_DR7\_BH\_fits.pro''). }

\begin{figure*}
  \centering
    \includegraphics[width=1.0\textwidth]{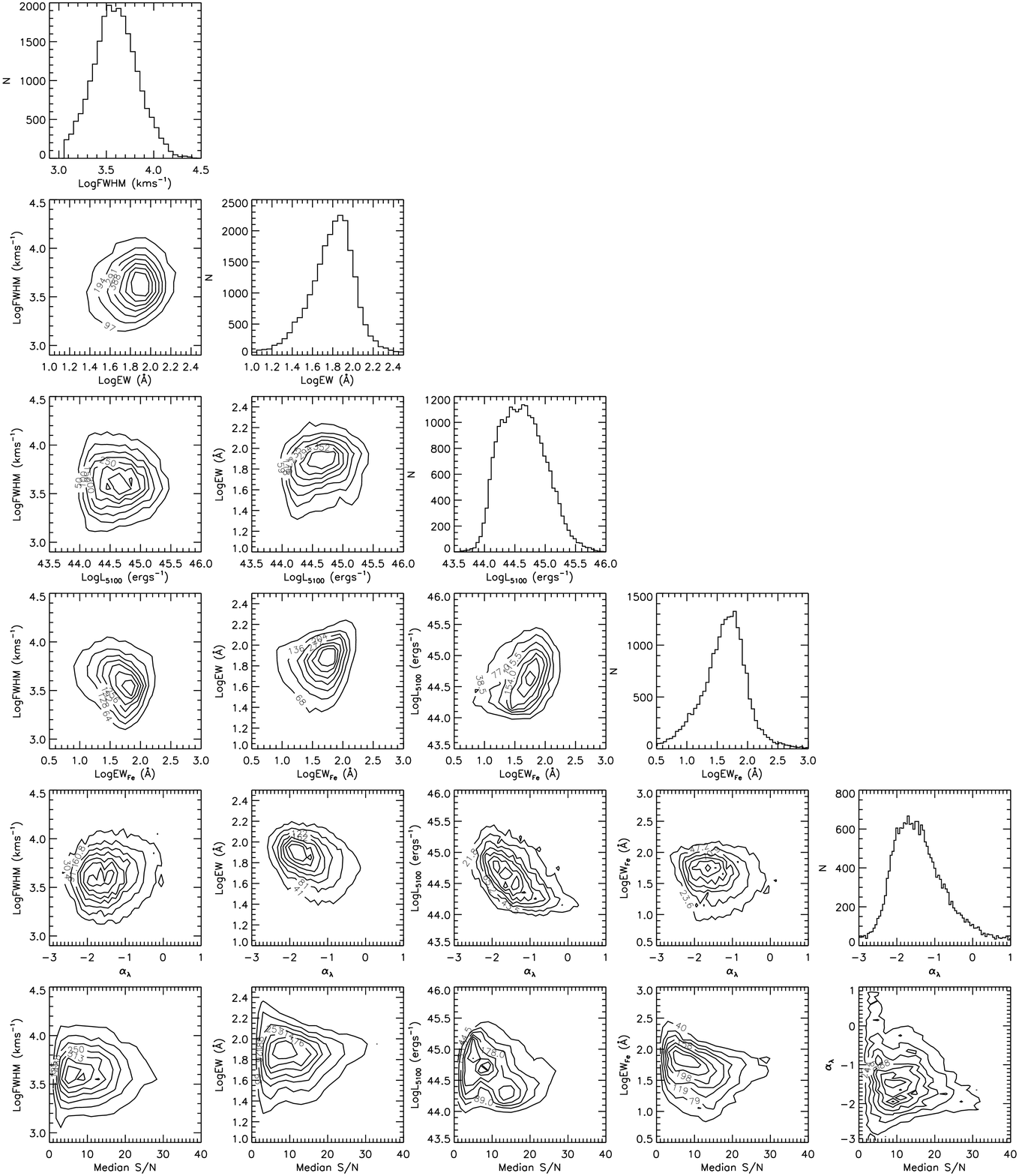}
    \caption{Statistical properties for \hbeta\ based on our spectral measurements (for all quasars). Contours are local point density contours
    estimated with a grid size of $\Delta=0.1$ on both axes ($\Delta=2$ on the median S/N axis). Contour levels are equally spaced linearly with
    their
    values marked for the outermost several contours. The strong anti-correlation between the power-law continuum slope and luminosity reflects the
    increasing host galaxy contamination towards fainter
    quasar luminosites (see \S\ref{subsec:host_contam} and Fig.\ \ref{fig:composite_spec}).}
    \label{fig:stat_hbeta}
\end{figure*}

\begin{figure*}
  \centering
    \includegraphics[width=1.\textwidth]{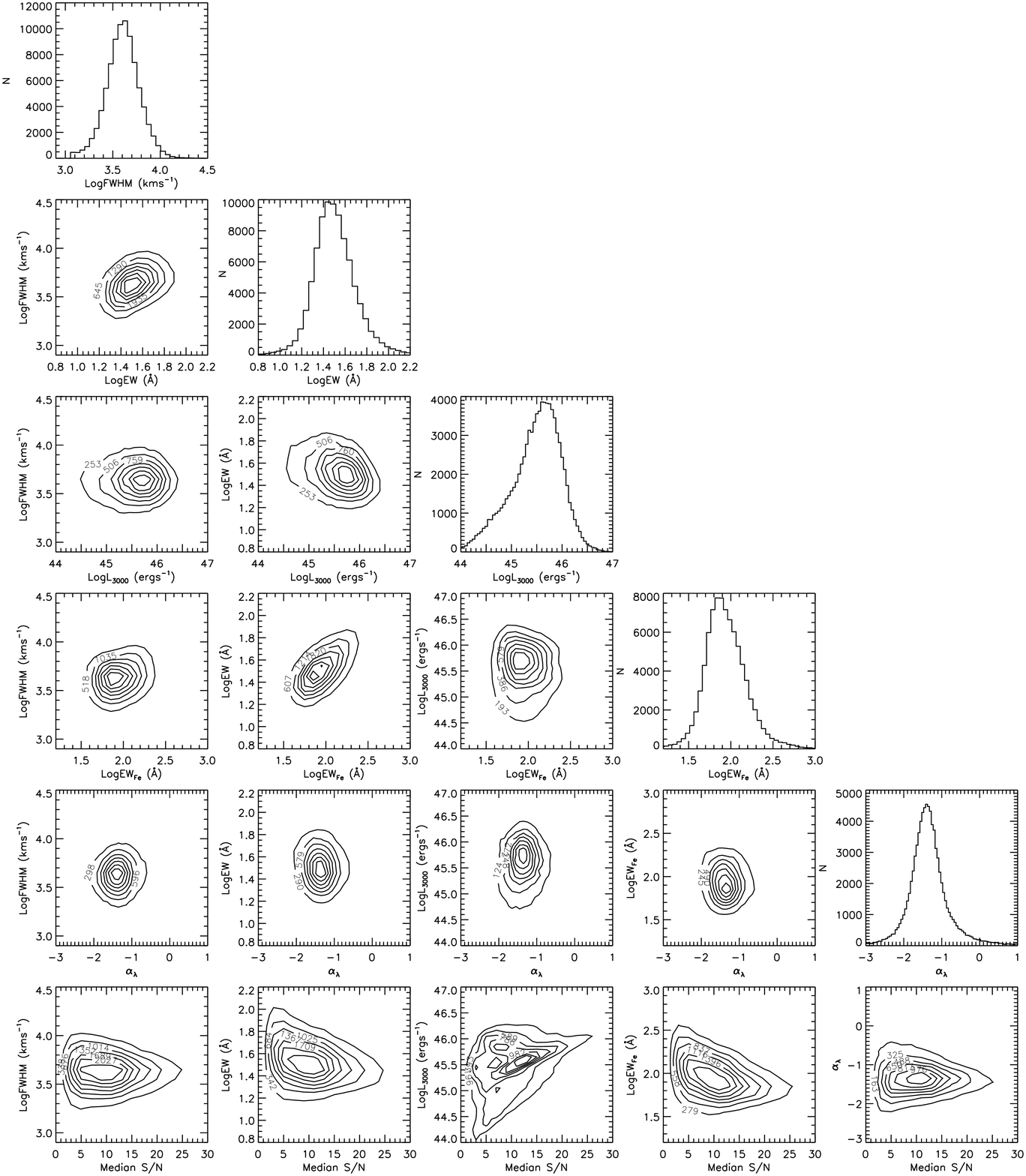}
    \caption{Statistical properties for \MgII\ based on our spectral measurements (for all quasars). Contours are local point density contours
    estimated with a grid size of $\Delta=0.1$ on both axes ($\Delta=2$ on the median S/N axis). Contour levels are equally spaced linearly with
    their
    values marked for the outermost several contours. There are several known correlations, such as the \MgII\ Baldwin effect
    \citep[e.g.,][]{Baldwin_1977,Croom_etal_2002}, and the correlation between FWHM and EW \citep[e.g.,][]{Dong_etal_2009b}. }
    \label{fig:stat_mgii}
\end{figure*}

\begin{figure*}
  \centering
    \includegraphics[width=1.\textwidth]{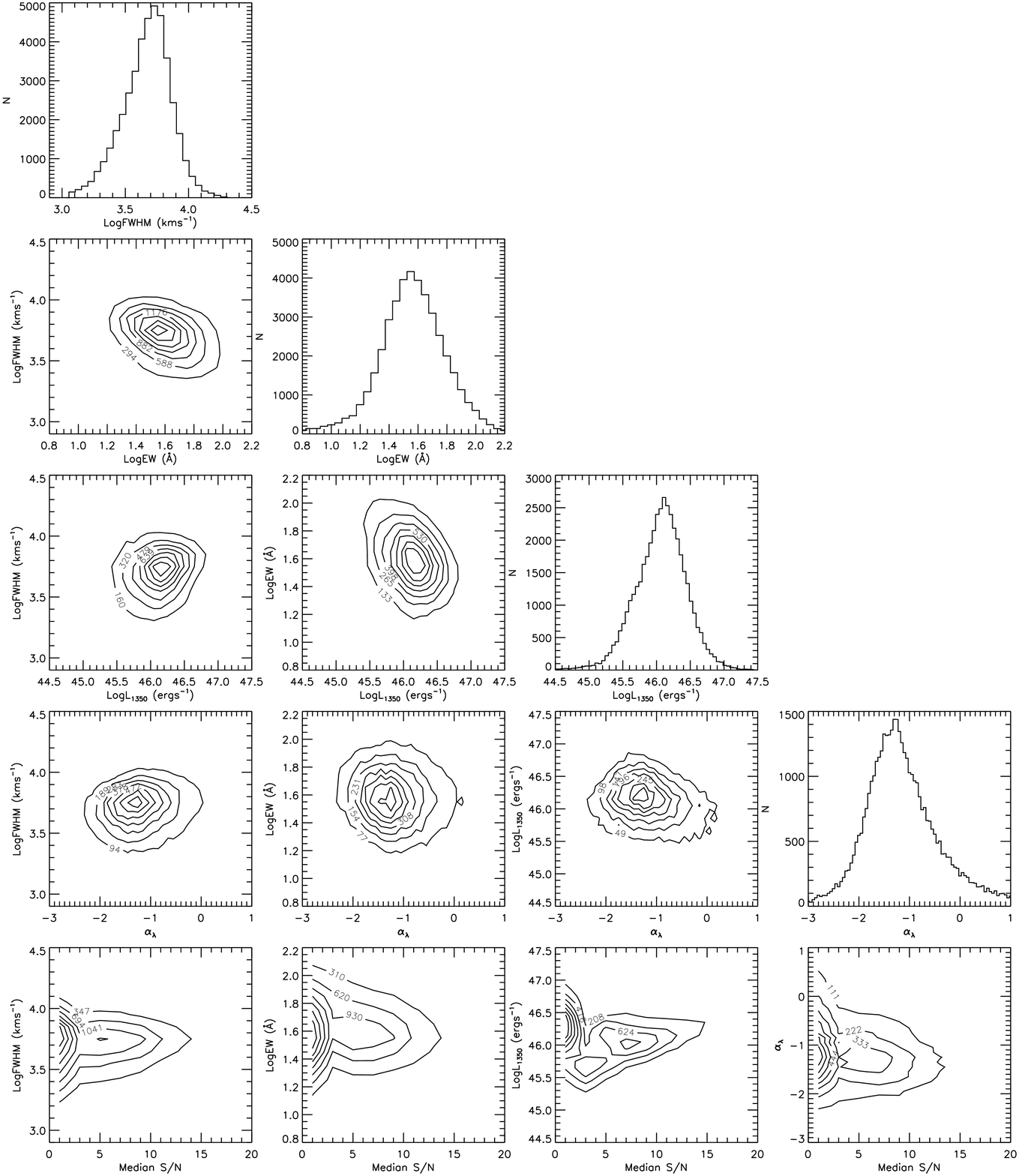}
    \caption{\footnotesize Statistical properties for \CIV\ based on our spectral measurements (for all quasars). Contours are local point density
    contours
    estimated with a grid size of $\Delta=0.1$ on both axes ($\Delta=2$ on the median S/N axis). Contour levels are equally spaced linearly with
    their
    values marked for the outermost several contours. There are several correlations involving EW, FWHM and
    luminosity that may be different manifestations of the same phenomenon \citep[e.g.,][]{Baldwin_1977,Richards_etal_2002b}. The clustering
    of a population of low S/N objects is caused by the quasar target selection, i.e., quasars are targeted to a fainter limiting magnitude at $z\ga
    2.9$ (see Fig.\ \ref{fig:mi_dist}). }
    \label{fig:stat_civ}
\end{figure*}

\subsection{Correlations between emission line properties}\label{subsec:corr}

One great virtue of the SDSS DR7 quasar survey is that it
provides unprecedented statistics for broad-line quasar
properties. To demonstrate this, Figs.\
\ref{fig:stat_hbeta}-\ref{fig:stat_civ} show some statistical
properties of quasars using our spectral measurements for
\hbeta, \MgII, and \CIV, respectively. These figures show the
typical values of these properties for SDSS quasars as a quick
reference. We do not show a similar figure for \halpha\ because
quasars with \halpha\ coverage represent only a tiny fraction
of the whole sample, and because host contamination is more
severe for these low-redshift quasars.

There are correlations among the properties shown in
Figs.\ \ref{fig:stat_hbeta}-\ref{fig:stat_civ}. Some of these correlations are not due to selection effects. For instance, the well-known Baldwin
effect \citep[][]{Baldwin_1977}, i.e., the anti-correlation between line EW and continuum luminosity, is clearly seen for \CIV\ and \MgII. There are
also strong correlations between EW and FWHM for \MgII\ \citep[e.g.,][]{Dong_etal_2009a} and \CIV, which are not due to any apparent selection
effects. The statistics of our catalog now allows in-depth investigations of these correlations when binning in different quantities such as redshift
or luminosity, and to probe the origins of these correlations. However, there are some apparent correlations which are likely due to selection
effects
inherent in a flux-limited sample, or host contamination. For instance, the apparent anti-correlation between $\alpha_\lambda$ and $\log L_{5100}$,
and the mild negative Baldwin effect below $\log L_{5100}\la 45$ for \hbeta\ seen in Fig.\ \ref{fig:stat_hbeta}, are most likely caused by increasing
host contamination towards fainter luminosities (see \S\ref{subsec:host_contam}). Moreover, the spectral quality (mainly S/N) has important effects
on
the measured quantities and may bias the measurements at the low S/N end. Thus one must take these issues into account when using the catalog to
study
correlations among various properties. The detailed investigations of various correlations will be presented elsewhere.

\begin{figure}
  \centering
    \includegraphics[width=0.45\textwidth]{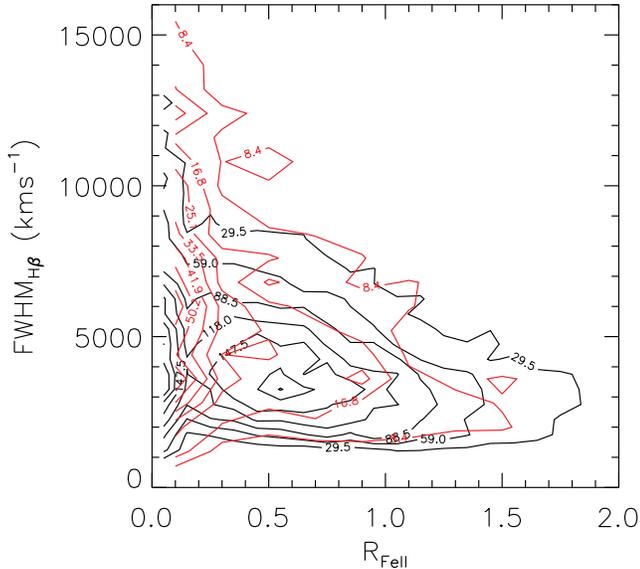}
    \caption{Distribution of quasars in the projected 4DE1 parameter space \citep[e.g.,][]{Sulentic_etal_2000}. Contours are
    local point density contours estimated with a grid size of $\Delta_x/\Delta_y=0.1/500$ (black) and $\Delta_x/\Delta_y=0.2/800$ (red), where the
    black contours are for all quasars and the red contours are for radio-loud ($R>10$) quasars only.}
    \label{fig:4DE1_Hbeta}
\end{figure}

\begin{figure}
  \centering
    \includegraphics[width=0.45\textwidth]{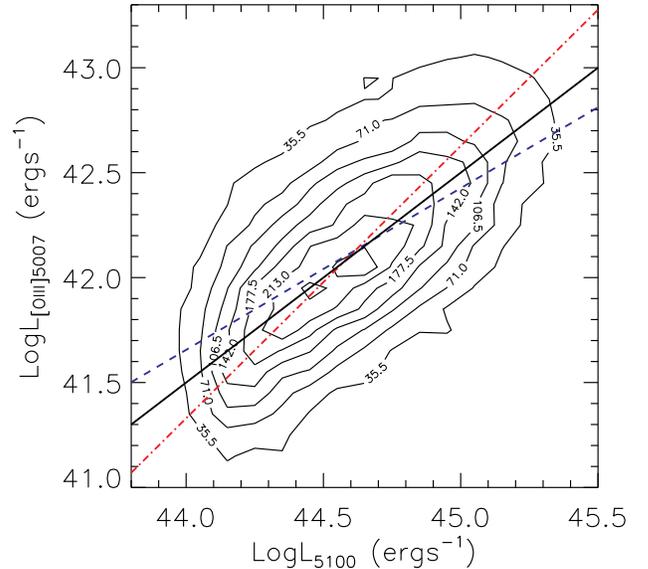}
    \caption{Correlation between $L_{\rm [OIII]5007}$ and $L_{5100}$. Contours are local point density contours estimated with a grid size of
    $\Delta=0.1$ on both axes. The dashed line is the linear regression fit treating $\log L_{5100}$ as the independent variable; the best-fit
    relation is $\log (L_{\rm [OIII]5007}/{\rm erg\,s^{-1}})=7.76+0.77\log (L_{5100}/{\rm erg\,s^{-1}})$ and the scatter around this relation is
    $\sim
    0.38$ dex. The dash-dotted line is the bisector linear regression fit; the best-fit relation is $\log (L_{\rm [OIII]5007}/{\rm
    erg\,s^{-1}})=-17.54+1.34\log (L_{5100}/{\rm erg\,s^{-1}})$ and the scatter around this relation is $\sim 0.4$ dex. The solid line is the mean
    linear relation described in Eqn.\ (\ref{eqn:Loiii_L5100}).
    }
    \label{fig:oiii_L5100}
\end{figure}

Fig.\ \ref{fig:4DE1_Hbeta} shows the so-called 4DE1 projection in
the \hbeta\ FWHM versus $R_{\rm FeII}={\rm
EW_{FeII4434-4684}/EW_{H\beta}}$ space
\citep[e.g.,][]{Sulentic_etal_2000,Sulentic_etal_2002,Zamfir_etal_2009}, which is an extension of the
eigenvector space for quasar properties suggested by \citet{Boroson_Green_1992}.
Objects with \hbeta\ FWHM $>4000\ {\rm km\,s^{-1}}$ (i.e.,
population ``B'' in the terminology of the 4DE1 parameter space
of Sulentic and collaborators) have a tendency to have weaker relative iron emission strength for larger FWHMs. The
black contours show the distribution of all quasars while the red
contours show the distribution of radio-loud ($R>10$) quasars. It
appears that the radio-loud contours are more vertically
elongated, broadly consistent with the phenomenological
classification scheme based on the 4DE1 parameter space
\citep[e.g.,][]{Sulentic_etal_2000,Sulentic_etal_2002,Zamfir_etal_2009}. The physics driving these characteristics in the parameter space is
currently
not clear, and deserves further study.

Fig.\ \ref{fig:oiii_L5100} shows the correlation between the
\OIIIb\ luminosity and the continuum luminosity at $5100$ \AA.
This correlation is usually used to estimate the bolometric
luminosity using the \OIIIb\ luminosity as a surrogate for type 2
quasars \citep[e.g,][]{Kauffmann_etal_2003,Zakamska_etal_2003,Heckman_etal_2004,
Reyes_etal_2008}. While the correlation is apparent, it has a large scatter, as noted in earlier studies
\citep[e.g.,][]{Heckman_etal_2004,Reyes_etal_2008}. The mean linear relation is:
\begin{equation}\label{eqn:Loiii_L5100}
\log L_{\rm [OIII]\lambda 5007}\approx \log L_{5100} -2.5\ ,
\end{equation}
with a scatter $\sim 0.35$ dex. Using the bolometric correction from \citet{Richards_etal_2006b}, a crude conversion between $L_{\rm [OIII]\lambda
5007}$ and the quasar bolometric luminosity is: $L_{\rm bol}\approx 3200L_{\rm [OIII]\lambda 5007}$.

%\begin{enumerate}
%
%\item {\bf Show the figures for \hbeta, \MgII, and \CIV\ for
%statistical properties, and discuss/compare with some claims such
%as Sulentic et al. for the 4DE1, Dong et al. for the iron
%correlation, as well as the Baldwin effect (Baldwin et al.,
%Richards et al., Croom et al.), and the scatter in line-width as
%function of luminosity (Fine et al.).}
%
%\item {\bf Correlations between \OIIIb\ luminosity and bolometric
%luminosity (cite papers by Kauffmann et al., Zakamska et al. and
%Reyes et al).}
%
%\end{enumerate}

\begin{figure*}
  \centering
    \includegraphics[width=0.9\textwidth]{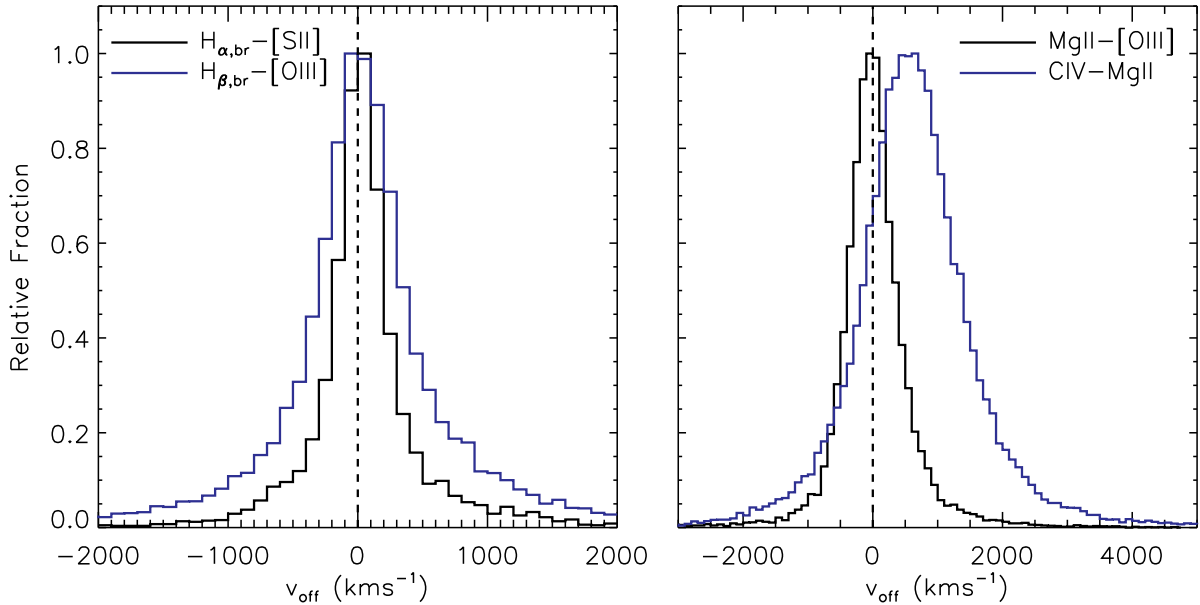}
    \caption{Velocity shifts between different pairs of lines. Positive values indicate blueshifts.}
    \label{fig:line_shift}
\end{figure*}

\subsection{Emission line shifts}

%\begin{enumerate}

%\item {\bf broad \hbeta\ and narrow line shifts. Note that many
%\OIIIb\ are blueshifted due to a blue
%wing\citep[e.g.,][]{Heckman_etal_1981,Komossa_etal_2008b}.}
%
%\item {\bf \MgII-\OIIIb\ shifts, \CIV-\MgII\ blue shifts.}

%\end{enumerate}

Fig.\ \ref{fig:line_shift} shows the distributions of velocity shifts between various emission lines. Recall that the velocity of the broad lines is
measured from the peak of the multiple-Gaussian fit. The left panel of Fig.\ \ref{fig:line_shift} shows the distributions of velocity shifts between
the broad Balmer lines and the narrow lines. The means of
these distributions are consistent with zero, hence
there is no offset in the mean between the broad and narrow Balmer lines
\citep[cf.,][]{Bonning_etal_2007}. We note that if we did not
account for the blue wings of the narrow \OIIIab\ lines during
spectral fitting, there would be a net redshift of the order of $\sim
100\ {\rm km\,s^{-1}}$ between the broad \hbeta\ line and \OIII, which is inconsistent with the results for \halpha\ versus
\SII. The right panel of Fig.\ \ref{fig:line_shift} shows the
velocity offsets between \MgII\ and \OIII\, and between \CIV\ and
\MgII. The \MgII\ line shows no mean offset from \OIII, while the
\CIV\ line shows a systematic blueshift of $\sim 600\ {\rm
km\,s^{-1}}$ with respect to \MgII\
\citep[e.g.,][]{Gaskell_1982,Tytler_Fan_1992,Richards_etal_2002b}; see \citet{Richards_etal_2011} for further discussion.

It is interesting to note that many of the objects in the wings of the velocity offset distributions of the broad Balmer lines vs the narrow lines
are
either strong disk-emitters
\citep[e.g.,][]{Chen_Halpern_1989,Eracleous_Halpern_1994,Strateva_etal_2003},
or have the broad component systematically offset from the narrow
line center; in other cases the apparent large shifts were
caused by poor fits to noisy spectra.

%{\bf Show some examples? Should I
%compile a list of objects that have real offsets? Should I compile
%a list of disk emitters?}

%{\bf Comparison between different virial estimators; discuss the
%caveats of the virial method (might want to cite some old papers
%by Debai and some reverberation mapping mass papers).}

%{\bf Need to work out a new version of \MgII\ virial estimator.
%The value of the $L-R$ slope, $a$, is essential. The goal is to
%match both the \hbeta\ estimates at low-$z$ (low-$L$) and the
%\CIV\ estimates at high-$z$ (high-$L$).}

\section{Summary}\label{sec:summary}

We have constructed an SDSS DR7 quasar catalog in which we tabulate various properties. In this catalog we compiled continuum and emission line
properties for \halpha, \hbeta, \MgII, and \CIV\ based on our spectral fits. We also included radio properties, and flagged quasars of special
interest, such as broad absorption line quasars and disk emitters. We also compiled virial BH mass estimates using these spectral measurements. This
catalog can be used to study correlations among properties of optically selected quasars, and the active black hole mass function in quasars (Shen et
al., in preparation).
We performed various tests and found that our automatic fitting procedure to emission lines performed reasonably well.
%Of course, for some particular purposes, more sophisticated measurements
% might be needed if the S/N allows.
However, one must take into account the possible effects of selection and S/N, as well as the systematics involved in
converting the measured quantities to derived quantities, when using these measurements to study quasar properties. In particular, we
{\em do not} encourage direct interpretations based on derived quantities (such as virial BH masses and bolometric luminosity) without accounting for
the difference between the estimated value and the true value for these quantities. Such direct interpretations will usually lead to biased or even
spurious
results.
%In particular, we re-emphasize the systematic issues with the current
% virial BH mass estimators, and urge caution upon usage of these mass estimates.

%One should also keep in mind that some derived quantities may have
%their own caveats. For instance, the virial BH mass estimates
%suffer from the statistical and systematic uncertainties of the
%virial estimators themselves. These caveats should be taken into
%account upon usage of these derived quantities.

We make this catalog publicly available online\footnote{
http://www.cfa.harvard.edu/$\sim$yshen/BH\_mass/dr7.htm}, where
we also provide supplemental materials (such as dereddened
spectra, QA plots, etc) and important future updates of this
compilation.

\acknowledgements

We thank the anonymous referee for constructive comments that improved the manuscript. YS acknowledges support from a Clay Postdoctoral Fellowship
through the Smithsonian Astrophysical Observatory (SAO). GTR acknowledges support from an Alfred P. Sloan Research Fellowship and NASA grant
07-ADP07-0035. MAS acknowledges the support of NSF grant AST-0707266. PBH is supported by NSERC. DPS acknowledges the support of NSF grant
AST-0607634.

Funding for the SDSS and SDSS-II has been provided by the Alfred
P. Sloan Foundation, the Participating Institutions, the National
Science Foundation, the U.S. Department of Energy, the National
Aeronautics and Space Administration, the Japanese Monbukagakusho,
the Max Planck Society, and the Higher Education Funding Council
for England. The SDSS Web Site is http://www.sdss.org/.

The SDSS is managed by the Astrophysical Research Consortium for
the Participating Institutions. The Participating Institutions are
the American Museum of Natural History, Astrophysical Institute
Potsdam, University of Basel, University of Cambridge, Case
Western Reserve University, University of Chicago, Drexel
University, Fermilab, the Institute for Advanced Study, the Japan
Participation Group, Johns Hopkins University, the Joint Institute
for Nuclear Astrophysics, the Kavli Institute for Particle
Astrophysics and Cosmology, the Korean Scientist Group, the
Chinese Academy of Sciences (LAMOST), Los Alamos National
Laboratory, the Max-Planck-Institute for Astronomy (MPIA), the
Max-Planck-Institute for Astrophysics (MPA), New Mexico State
University, Ohio State University, University of Pittsburgh,
University of Portsmouth, Princeton University, the United States
Naval Observatory, and the University of Washington.

Facilities: Sloan

%\clearpage
%\bibliography{D:/Research/bib/smbhrefs}

%\clearpage

%\begin{deluxetable*}{ccc}\tabletypesize{\tiny}
%\tablecolumns{3}\tablewidth{0.8\textwidth} \tablecaption{Catalog
%Format} \tablehead{Column & Format & Description} \startdata
%\tiny
\begin{longtable}{ccc}
\caption[notes]{FITS Catalog Format}\label{tab:format}\\
\hline \hline \\[-2ex]
   \multicolumn{1}{c}{\textbf{Column}} &
   \multicolumn{1}{c}{\textbf{Format}} &
   \multicolumn{1}{c}{\textbf{Description}} \\[0.5ex] \hline
   \\[-1.8ex]
\endfirsthead
%This is the header for the remaining page(s) of the table...
\multicolumn{3}{c}{{\tablename} \thetable{} -- Continued} \\[0.5ex]
  \hline \hline \\[-2ex]
  \multicolumn{1}{c}{\textbf{Column}} &
  \multicolumn{1}{c}{\textbf{Format}} &
  \multicolumn{1}{c}{\textbf{Description}} \\[0.5ex] \hline
  \\[-1.8ex]
\endhead
%This is the footer for all pages except the last page of the table...
  \hline
  \multicolumn{3}{l}{{Continued on Next Page\ldots}} \\
\endfoot
%This is the footer for the last page of the table...
  \\[-1.8ex] \hline \hline
\endlastfoot
1\dotfill  & STRING   & SDSS DR7 designation hhmmss.ss$+$ddmmss.s (J2000.0)\\
2\dotfill   & DOUBLE & Right ascension in decimal degrees (J2000.0) \\
3\dotfill   & DOUBLE & Declination in decimal degrees (J2000.0) \\
4\dotfill   & DOUBLE   & Redshift \\
5\dotfill   & LONG       & Spectroscopic plate number \\
6\dotfill   & LONG       & Spectroscopic fiber number \\
7\dotfill   & LONG       & MJD of spectroscopic observation\\
8\dotfill   & LONG     & Target selection flag (TARGET version)\\
9\dotfill   & LONG       & Number of spectroscopic observations\\
10\dotfill & LONG       & Uniform selection flag\\
11\dotfill & DOUBLE   & $M_i (z=2)$ [$h=0.7$, $\Omega_0=0.3$,
$\Omega_\Lambda=0.7$, $K$-corrected to $z=2$, following \citet{Richards_etal_2006a}]\\
12\dotfill & DOUBLE  & Bolometric luminosity [$\log (L_{\rm bol}/\rm erg
s^{-1})$]\\
13\dotfill & DOUBLE  & Uncertainty in $\log L_{\rm bol}$\\
14\dotfill & LONG      & BAL flag (0=nonBALQSO or no wavelength coverage; 1=\CIV\ BALQSO; 2=\MgII\ BALQSO; 3=both 1 and 2)\\
15\dotfill & LONG      & FIRST radio flag (-1=not in FIRST footprint; 0=FIRST undetected; 1=core-dominant; 2=lobe-dominant)\\
16\dotfill & DOUBLE    & Observed radio flux density at rest-frame 6 cm $f_{\rm\nu, 6cm}$ [mJy]\\
17\dotfill & DOUBLE    & Observed optical flux density at rest-frame 2500\AA\ [$\log (f_{\nu,2500}/{\rm ergs^{-1}cm^{-2}Hz^{-1}})$]\\
18\dotfill & DOUBLE   & Radio loudness $R\equiv f_{\rm\nu, 6cm}/f_{\nu,2500}$\\
19\dotfill & DOUBLE   & Monochromatic luminosity at 5100\AA\ [$\log (L_{\rm 5100}/\rm ergs^{-1})$]\\
20\dotfill & DOUBLE   & Uncertainty in $\log L_{\rm 5100}$\\
21\dotfill & DOUBLE   & Monochromatic luminosity at 3000\AA\ [$\log (L_{\rm 3000}/\rm ergs^{-1})$]\\
22\dotfill & DOUBLE   & Uncertainty in $\log L_{\rm 3000}$\\
23\dotfill & DOUBLE   & Monochromatic luminosity at 1350\AA\ [$\log (L_{\rm 1350}/\rm ergs^{-1})$]\\
24\dotfill & DOUBLE   & Uncertainty in $\log L_{\rm 1350}$\\
25\dotfill & DOUBLE   & Line luminosity of broad \halpha\ [$\log (L/\rm ergs^{-1})$]\\
26\dotfill & DOUBLE   & Uncertainty in $\log L_{{\rm H\alpha,broad}}$\\
27\dotfill & DOUBLE     & FWHM of broad \halpha\ (${\rm kms^{-1}}$)\\
28\dotfill & DOUBLE     & Uncertainty in the broad \halpha\ FWHM\\
29\dotfill & DOUBLE   & Restframe equivalent width of broad \halpha\ (\AA)\\
30\dotfill & DOUBLE   & Uncertainty in ${\rm EW_{H\alpha,broad}}$\\
31\dotfill & DOUBLE   & Line luminosity of narrow \halpha\ [$\log (L/\rm ergs^{-1})$]\\
32\dotfill & DOUBLE   & Uncertainty in $\log L_{{\rm H\alpha,narrow}}$\\
33\dotfill & DOUBLE     & FWHM of narrow \halpha\ (${\rm kms^{-1}}$)\\
34\dotfill & DOUBLE     & Uncertainty in the narrow \halpha\ FWHM\\
35\dotfill & DOUBLE   & Restframe equivalent width of narrow \halpha\ (\AA)\\
36\dotfill & DOUBLE   & Uncertainty in ${\rm EW_{H\alpha,narrow}}$\\
37\dotfill & DOUBLE   & Line luminosity of \NIIb\ [$\log (L/\rm ergs^{-1})$]\\
38\dotfill & DOUBLE   & Uncertainty in $\log L_{{\rm [NII]6584}}$\\
39\dotfill & DOUBLE   & Restframe equivalent width of \NIIb\ (\AA)\\
40\dotfill & DOUBLE   & Uncertainty in ${\rm EW_{[NII]6584}}$\\
41\dotfill & DOUBLE   & Line luminosity of \SIIa\ [$\log (L/\rm ergs^{-1})$]\\
42\dotfill & DOUBLE   & Uncertainty in $\log L_{\rm [SII]6717}$\\
43\dotfill & DOUBLE   & Restframe equivalent width of \SIIa\ (\AA)\\
44\dotfill & DOUBLE   & Uncertainty in ${\rm EW_{[SII]6717}}$\\
45\dotfill & DOUBLE   & Line luminosity of \SIIb\ [$\log (L/\rm ergs^{-1})$]\\
46\dotfill & DOUBLE   & Uncertainty in $\log L_{\rm [SII]6731}$\\
47\dotfill & DOUBLE   & Restframe equivalent width of \SIIb\ (\AA)\\
48\dotfill & DOUBLE   & Uncertainty in ${\rm EW_{[SII]6731}}$\\
49\dotfill & DOUBLE   & Restframe equivalent width of Fe within 6000-6500\AA\ (\AA)\\
50\dotfill & DOUBLE   & Uncertainty in ${\rm EW_{Fe,H\alpha}}$\\
51\dotfill & DOUBLE   & Power-law slope for the continuum fit for \halpha\ \\
52\dotfill & DOUBLE   & Uncertainty in $\alpha_{\rm H\alpha}$\\
53\dotfill & LONG       & Number of good pixels for the restframe 6400-6765\AA\ region\\
54\dotfill & DOUBLE   & Median S/N per pixel for  the restframe 6400-6765\AA\ region\\
55\dotfill & DOUBLE   & Reduced $\chi^2$ for the \halpha\ line fit; -1 if not fitted\\
56\dotfill & DOUBLE   & Line luminosity of broad \hbeta\ [$\log (L/\rm ergs^{-1})$]\\
57\dotfill & DOUBLE   & Uncertainty in $\log L_{\rm H\beta,broad}$\\
58\dotfill & DOUBLE     & FWHM of broad \hbeta\ (${\rm kms^{-1}}$)\\
59\dotfill & DOUBLE     & Uncertainty in the broad \hbeta\ FWHM\\
60\dotfill & DOUBLE   & Restframe equivalent width of broad \hbeta\ (\AA)\\
61\dotfill & DOUBLE   & Uncertainty in ${\rm EW_{H\beta,broad}}$\\
62\dotfill & DOUBLE   & Line luminosity of narrow \hbeta\ [$\log (L/\rm ergs^{-1})$]\\
63\dotfill & DOUBLE   & Uncertainty in $\log L_{\rm H\beta,narrow}$\\
64\dotfill & DOUBLE     & FWHM of narrow \hbeta\ (${\rm kms^{-1}}$)\\
65\dotfill & DOUBLE     & Uncertainty in the narrow \hbeta\ FWHM\\
66\dotfill & DOUBLE   & Restframe equivalent width of narrow \hbeta\ (\AA)\\
67\dotfill & DOUBLE   & Uncertainty in ${\rm EW_{H\beta,narrow}}$\\
68\dotfill & DOUBLE     & FWHM of broad \hbeta\ using a single Gaussian fit (${\rm kms^{-1}}$)\\
69\dotfill & DOUBLE   & Line luminosity of \OIIIa\ [$\log (L/\rm ergs^{-1})$]\\
70\dotfill & DOUBLE   & Uncertainty in $\log L_{\rm [OIII]4959}$\\
71\dotfill & DOUBLE   & Restframe equivalent width of \OIIIa\ (\AA)\\
72\dotfill & DOUBLE   & Uncertainty in ${\rm EW_{[OIII]4959}}$\\
73\dotfill & DOUBLE   & Line luminosity of \OIIIb\ [$\log (L/\rm ergs^{-1})$]\\
74\dotfill & DOUBLE   & Uncertainty in $\log L_{\rm [OIII]5007}$\\
75\dotfill & DOUBLE   & Restframe equivalent width of \OIIIb\ (\AA)\\
76\dotfill & DOUBLE   & Uncertainty in ${\rm EW_{[OIII]5007}}$\\
77\dotfill & DOUBLE   & Restframe equivalent width of Fe within 4435-4685\AA\ (\AA)\\
78\dotfill & DOUBLE   & Uncertainty in ${\rm EW_{Fe,H\beta}}$\\
79\dotfill & DOUBLE   & Power-law slope for the continuum fit for \hbeta\ \\
80\dotfill & DOUBLE   & Uncertainty in $\alpha_{\rm H\beta}$\\
81\dotfill & LONG       & Number of good pixels for the restframe 4750-4950\AA\ region\\
82\dotfill & DOUBLE   & Median S/N per pixel for  the restframe 4750-4950\AA\ region\\
83\dotfill & DOUBLE   & Reduced $\chi^2$ for the \hbeta\ line fit; -1 if not fitted\\
84\dotfill & DOUBLE   & Line luminosity of the whole \MgII\ [$\log (L/\rm ergs^{-1})$]\\
85\dotfill & DOUBLE   & Uncertainty in $\log L_{\rm MgII,whole}$\\
86\dotfill & DOUBLE     & FWHM of the whole \MgII\ (${\rm kms^{-1}}$)\\
87\dotfill & DOUBLE     & Uncertainty in the whole \MgII\ FWHM\\
88\dotfill & DOUBLE   & Restframe equivalent width of the whole \MgII\ (\AA)\\
89\dotfill & DOUBLE   & Uncertainty in ${\rm EW_{MgII,whole}}$
\\
90\dotfill & DOUBLE   & Line luminosity of broad \MgII\ [$\log (L/\rm ergs^{-1})$]\\
91\dotfill & DOUBLE   & Uncertainty in $\log L_{\rm MgII,broad}$\\
92\dotfill & DOUBLE       & FWHM of broad \MgII\ (${\rm kms^{-1}}$)\\
93\dotfill & DOUBLE     & Uncertainty in the broad \MgII\ FWHM\\
94\dotfill & DOUBLE   & Restframe equivalent width of broad \MgII\ (\AA)\\
95\dotfill & DOUBLE   & Uncertainty in ${\rm EW_{MgII,broad}}$\\
96\dotfill & DOUBLE       & FWHM of broad \MgII\ using a single Gaussian fit (${\rm kms^{-1}}$)\\
97\dotfill & DOUBLE   & Restframe equivalent width of Fe within 2200-3090\AA\ (\AA)\\
98\dotfill & DOUBLE   & Uncertainty in ${\rm EW_{Fe,MgII}}$\\
99\dotfill & DOUBLE   & Power-law slope for the continuum fit for \MgII\ \\
100\dotfill & DOUBLE  & Uncertainty in $\alpha_{\rm MgII}$\\
101\dotfill & LONG       & Number of good pixels for the restframe 2700-2900\AA\ region\\
102\dotfill & DOUBLE   & Median S/N per pixel for  the restframe 2700-2900\AA\ region\\
103\dotfill & DOUBLE   & Reduced $\chi^2$ for the \MgII\ line fit; -1 if not fitted\\
104\dotfill & DOUBLE   & Line luminosity of the whole \CIV\ [$\log (L/\rm ergs^{-1})$]\\
105\dotfill & DOUBLE   & Uncertainty in $\log L_{\rm CIV}$\\
106\dotfill & DOUBLE       & FWHM of the whole \CIV\ (${\rm kms^{-1}}$)\\
107\dotfill & DOUBLE     & Uncertainty in the \CIV\ FWHM\\
108\dotfill & DOUBLE   & Restframe equivalent width of the whole \CIV\ (\AA)\\
109\dotfill & DOUBLE   & Uncertainty in ${\rm EW_{CIV}}$\\
110\dotfill & DOUBLE   & Power-law slope for the continuum fit for \CIV\ \\
111\dotfill & DOUBLE   & Uncertainty in $\alpha_{\rm CIV}$\\
112\dotfill & LONG       & Number of good pixels for the restframe 1500-1600\AA\ region\\
113\dotfill & DOUBLE   & Median S/N per pixel for  the restframe 1500-1600\AA\ region\\
114\dotfill & DOUBLE   & Reduced $\chi^2$ for the \CIV\ fit; -1 if not fitted\\
115\dotfill & DOUBLE       & Velocity shift of broad \halpha\ (${\rm kms^{-1}}$); 3d5 if not measurable\\
116\dotfill & DOUBLE      & Uncertainty in $V_{\rm H\alpha, broad}$\\
117\dotfill & DOUBLE       & Velocity shift of narrow \halpha\ (${\rm kms^{-1}}$); 3d5 if not measurable\\
118\dotfill & DOUBLE      & Uncertainty in $V_{\rm H\alpha, narrow}$\\
119\dotfill & DOUBLE       & Velocity shift of broad \hbeta\ (${\rm kms^{-1}}$); 3d5 if not measurable\\
120\dotfill & DOUBLE      & Uncertainty in $V_{\rm H\beta, broad}$\\
121\dotfill & DOUBLE       & Velocity shift of narrow \hbeta\ (${\rm kms^{-1}}$); 3d5 if not measurable\\
122\dotfill & DOUBLE      & Uncertainty in $V_{\rm H\beta, narrow}$\\
123\dotfill & DOUBLE       & Velocity shift of broad \MgII\ (${\rm kms^{-1}}$); 3d5 if not measurable\\
124\dotfill & DOUBLE      & Uncertainty in $V_{\rm MgII, broad}$\\
125\dotfill & DOUBLE       & Velocity shift of \CIV\ (${\rm kms^{-1}}$); 3d5 if not measurable\\
126\dotfill & DOUBLE      & Uncertainty in $V_{\rm CIV}$\\
127\dotfill & DOUBLE   & Virial BH mass based on \hbeta\ [MD04, $\log (M_{\rm BH,vir}/M_\odot$)]\\
128\dotfill & DOUBLE   & Measurement uncertainty in $\log M_{\rm BH,vir}$ (\hbeta, MD04)\\
129\dotfill & DOUBLE   & Virial BH mass based on \hbeta\ [VP06, $\log (M_{\rm BH,vir}/M_\odot$)]\\
130\dotfill & DOUBLE   & Measurement uncertainty in $\log M_{\rm BH,vir}$ (\hbeta, VP06)\\
131\dotfill & DOUBLE   & Virial BH mass based on \MgII\ [MD04, $\log (M_{\rm BH,vir}/M_\odot$)]\\
132\dotfill & DOUBLE   & Measurement uncertainty in $\log M_{\rm BH,vir}$ (\MgII, MD04)\\
133\dotfill & DOUBLE   & Virial BH mass based on \MgII\ [VO09, $\log (M_{\rm BH,vir}/M_\odot$)]\\
134\dotfill & DOUBLE   & Measurement uncertainty in $\log M_{\rm BH,vir}$ (\MgII, VO09)\\
135\dotfill & DOUBLE   & Virial BH mass based on \MgII\ [S10, $\log (M_{\rm BH,vir}/M_\odot$)]\\
136\dotfill & DOUBLE   & Measurement uncertainty in $\log M_{\rm BH,vir}$ (\MgII, S10)\\
137\dotfill & DOUBLE   & Virial BH mass based on \CIV\ [VP06, $\log (M_{\rm BH,vir}/M_\odot$)]\\
138\dotfill & DOUBLE   & Measurement uncertainty in $\log M_{\rm BH,vir}$ (\CIV, VP06)\\
139\dotfill & DOUBLE   & The adopted fiducial virial BH mass [$\log (M_{\rm BH,vir}/M_\odot$)]\\
140\dotfill & DOUBLE   & Uncertainty in the fiducial virial BH mass (measurement uncertainty only)\\
141\dotfill & DOUBLE   & Eddington ratio based on the fiducial virial BH mass [$\log (L_{\rm bol}/L_{\rm Edd})$]\\
142\dotfill & LONG       & Special interest flag \\
%\enddata
%\end{deluxetable*}
\end{longtable}\scriptsize
\tablecomments{(1) Objects are in the same order as
in the DR7 quasar catalog \citep{Schneider_etal_2010}; (2)
$K-$corrections are the same as in \citet{Richards_etal_2006a}; (3)
Bolometric luminosities computed using bolometric corrections in
\citet{Richards_etal_2006b} using one of the $5100$\AA, $3000$\AA, or
$1350$\AA\ monochromatic luminosities depending on redshift; (4) Uncertainties are measurement errors only; (5) Unless otherwise stated, null value
(indicating unmeasurable) is zero for a quantity and $-1$ for its associated error.}
\normalsize

\end{document}